\documentclass[prb,twocolumn,showpacs,preprintnumbers]{revtex4-2}

\usepackage[utf8]{inputenc} 
\usepackage{graphicx}
\usepackage{ifthen}
\usepackage{ifpdf}
\usepackage{color}
\definecolor{red}{rgb}{1.,0.,0.}
\definecolor{blue}{rgb}{0.,0.,1.}

\usepackage{latexsym}
\usepackage{amsmath}
\usepackage{bm}
\usepackage{bbm} 
\newcommand{\half}{{\textstyle\frac{1}{2}}}

\newcommand{\beq}[1]{\begin{eqnarray}\ifthenelse{#1=-1}{\nonumber}
{\ifthenelse{#1=0}{}{\label{e#1}}}}
\newcommand{\eeq}{\end{eqnarray}}
\newcommand{\be}{\begin{equation}}
\newcommand{\ee}{\end{equation}}

\newcommand{\bea}{\begin{eqnarray}}
\newcommand{\eea}{\end{eqnarray}}

\newcommand{\hide}[1]{}

\usepackage{lineno}

\begin{document}

\title{NMR of a single nuclear spin detected by a scanning tunneling microscope}
\author{Yishay Manassen$^1$, Michael Averbukh$^1$, Zion Hazan$^1$, Yahel Tzuriel$^1$, Pino Boscolo$^2$, Alexsander Shnirman$^3$ and Baruch Horovitz$^{1}$}
\affiliation{$^1$Department of Physics, Ben Gurion University of the Negev, Beer Sheva 84105, Israel\\
 $^2$Gruppo Techniche Avanzate, Via Vergerio 1, 34138 Trieste, Italy\\
 $^3$Institut für Theorie der Kondensierten Materie, Karlsruhe Institute of Technology, D-76131 Karlsruhe, Germany}
\begin{abstract}
We demonstrate ionization of a molecule with the bias voltage of a Scanning Tunnelling Microscope (STM) resulting in a coexistence of a neutral and ionic molecules, i.e.  radical (paramagnetic) and non-radical (diamagnetic) states. This coexistence may be facilitated by a periodic switching between two bias voltages.
 The precession of the nucleus in the diamagnetic state modulates the nuclear polarization as well as the hyperfine transitions as seen in electron spin resonance (ESR). We analyze the power spectrum of a selected hyperfine intensity and obtain the nuclear magnetic resonance (NMR) spectrum. We have observed this phenomenon in three types of molecules: TEMPO, toluene and triphenylphosphine, showing NMR of  $^{14}$N, $^{13}$C, $^{31}$P and $^1$H nuclei. The spectra are detailed and show signatures of the chemical environment, i.e. chemical shifts. A master equation including off-diagonal hyperfine interactions accounts for these observations.
\end{abstract}
\maketitle

The detection of Nuclear Magnetic Resonance (NMR) of individual molecules is an outstanding challenge. It is essential for chemical analysis, for quantum information devices and as a significant improvement of medical NMR. Furthermore, a high resolution NMR can identify the chemical environment of a given nucleus. Previous local NMR probes were attempted by magnetic resonance force microscopy \cite{mamin,degen} that have reached a resolution of 10nm or by NV centers \cite{muller,pfender} that, however, lack the flexibility of scanning.

In the present work we measure electron-spin-resonance (ESR) by the technique of ESR-STM \cite{manassen1,manassen2,manassen3} to probe via the hypefine coupling individual nuclear spins. In contrast with other ESR-STM methods \cite{baumann,willke} in our scheme we do not need either a polarized STM tip, high magnetic fields, low temperatures or even external rf fields, thus we have the advantage of simplicity.
\\

{\bf Principles and Outline}

The present work can be applied to either radicals or non-radical molecules, The basic idea is to ionize the molecule using the STM bias creating a coexistence of neutral and ionized states. In the radical state the hyperfine tensor couples the electron and nuclear spins, $\bm\sigma$ and $\bm\tau$ respectively. Consider an ESR frequency $\nu=g\mu_B B$, where $g$ is the g-factor, $\mu_B$ is the Bohr magneton and $B$ is the external magnetic field in the $z$ direction. The dominant hyperfine terms are $a\sigma_z\tau_z$ and an off-diagonal $d\sigma_z\tau_y$, present for a general molecular orientation, assuming $\nu\gg a,d$. When the eigenvalue of $a\tau_z+d\tau_y$ is positive (negative) the hyperfine transition is at $\nu+\sqrt{a^2+d^2}$ ($\nu-\sqrt{a^2+d^2}$). Thus, as $\bm\tau$ rotates, e.g. freely in the non-radical state, the intensity of a given hyperfine resonance is oscillating in time, hence detectable by ESR-STM. The challenge is then, both experimentally and theoretically, to create conditions where the free nucleus encodes its phase on the ESR spectrum.

In our experiments we probe various nuclei such as $^{14}$N, $^{13}$C, $^{31}$P and $^1$H . The STM bias voltage is modulated in time so as to enhance ionization, though we find that this modulation is not essential. We then record a given hyperfine intensity during a sequence of dwell times, each one is short relative to the nuclear period. Finally we identify the power spectrum of this sequence as our NMR spectrum. We find that in 1sec we can take an NMR spectrum of each pixel in an STM image (see Supplementary Information (SM) \cite{SM}). Our experiments are then a proof of concept, providing a powerful yet simple technique for detection of single nucleus NMR. Furthermore, the NMR spectrum with high frequency resolution provides chemical shifts \cite{tolman}, spin-spin (dipolar) interactions, hence the chemical environment of the nucleus.

We demonstrate our technique on three types of molecules: (2,2,6,6-Tetramethylpiperidin-1-yl)oxyl (TEMPO), which is a radical at zero bias and becomes a non-radical in its ionized state, toluene and triphenylphosphine both molecules are non-radicals at zero bias and become radicals in the ionized state, see structures and STM images in  Fig. 1. We note also recent experiments \cite{hazan} on $^{60}$C which is non-radical at zero bias, becoming a radical at finite STM bias as seen by its ESR-STM spectrum.
\\

\begin{figure}[t]
\textit{}\includegraphics*[height=.35\columnwidth]{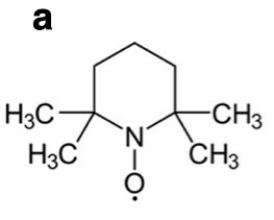}\hspace{2mm}
\includegraphics*[height=.35\columnwidth]{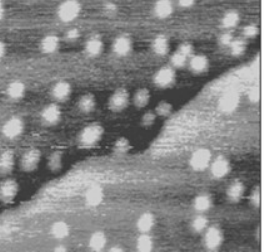}

\vspace{.5cm}
\includegraphics*[height=.35\columnwidth]{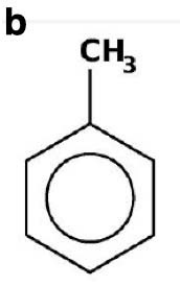}\hspace{2cm}
\includegraphics*[height=.35\columnwidth]{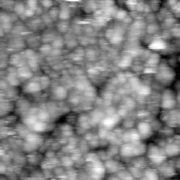}

\vspace{.5cm}
\includegraphics*[height=.35\columnwidth]{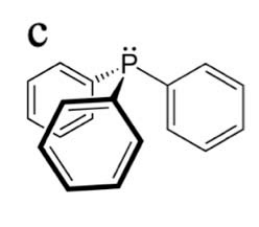}\hspace{4mm}
\includegraphics*[height=.35\columnwidth]{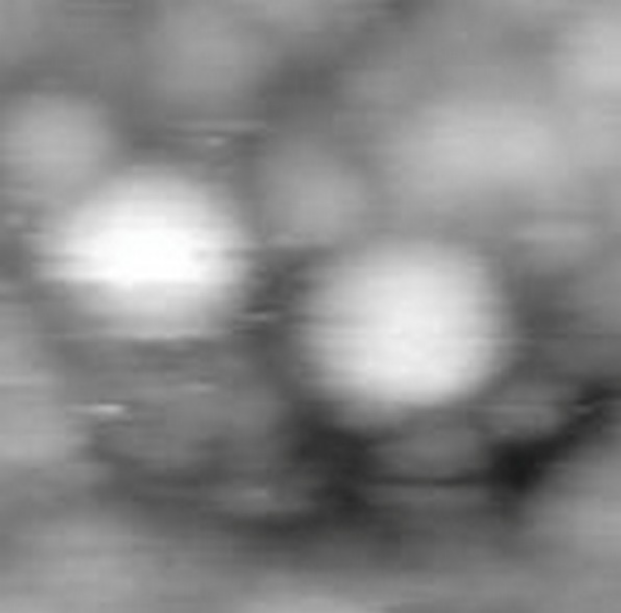}
\caption{ {\bf Molecular structures and STM images}. Left: atomic structure of molecules, right: their STM image (20x20nm$^2$ for a,b, 10x10nm$^2$ for c). STM current is I=0.1nA and voltage 1V. (a) TEMPO on Au(111) covered with graphene oxide monolayer, (b) toluene on Au(111), (c) triphenylphosphine on Au(111).}
\end{figure}

{\bf NMR-STM Experiments}

 Consider first our NMR-STM experiment on TEMPO molecules; we note our earlier ESR-STM data for various TEMPO configurations \cite{dimer}. We have chosen the ESR intensity at 760 MHz and recorded its value during 0.25 $\mu$sec (dwell time $T_d$) by a fast scope (Rhode Schwarz), taking $10^6$ values, i.e. total of 0.25 sec. This set of ESR intensities forms a sequence $h(t)$ (proportional to the ESR intensity in dBm units) at times $t$ separated by $T_d$. The corresponding Fourier $|h(\nu)|^2$ representing the NMR spectrum is then plotted. We have taken 50 spectra within 5 min forming one group, then searched the STM image for another molecule and repeated this procedure, thus generating 101 groups. The nominal magnetic field at the beginning of the experiment was 233 G though it was not monitored during the experiment. Fig. 2a shows the full range of the spectrum, summing on the 50 spectra of one group. A strong line is visible at 0.0785 MHz and is present in about 1/3 of the groups. We assign this line to $^{14}$N, with the standard $^{14}$N gyromagnetic factor 3.077 MHz/T this line corresponds to a field of 255 G. A high resolution plot is shown in Fig. 2b of the sum on 50 spectra as well as that of one individual spectrum. The linewidth is $\sim$100 Hz, as common in molecules with $^{14}$N \cite{witanowski}. The individual spectrum exhibits considerable structure, consistent with known chemical shifts \cite{witanowski}. There are two weaker lines at 0.2303 MHz and at 1.544 MHz that have a narrow linewidth of $\sim$ 20 Hz, we consider these lines as due to extrinsic noise.

\begin{figure}[t]
\includegraphics*[height=.4\columnwidth]{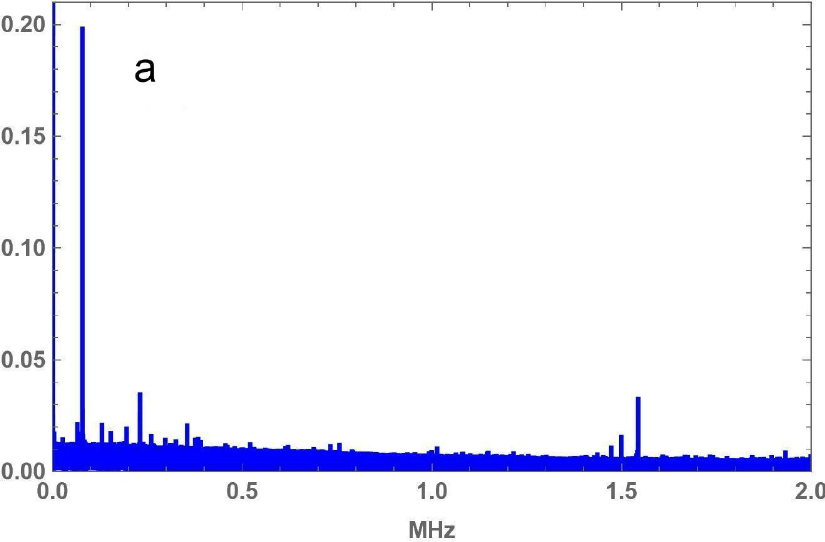}
\includegraphics*[height=.4\columnwidth]{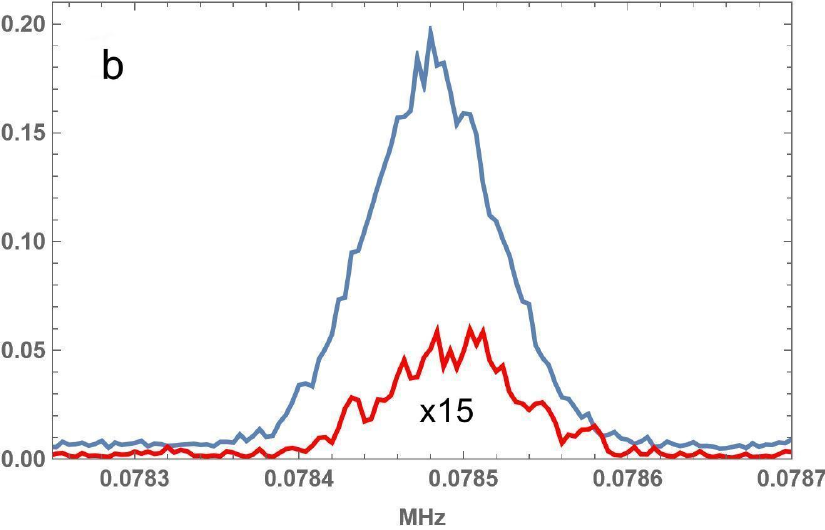}
\caption{{\bf Data on TEMPO}. (a,b) NMR data using the fast scope method. Monitored ESR frequency is 760 MHz, voltage is modulated between 0.2V and 3.8V at 15 MHz, tunneling current is I=0.5 nA, nominal B=233 G, $T_d=0.25\mu$sec. Data is summed over 50 consecutive spectra while (b) shows also a single spectrum (lower red line, enhanced by x15). }
\end{figure}

Our next set of experiments shows $^1$H nuclei in toluene, Fig. 3. ESR-STM data is shown in Fig. S2 \cite{SM}. This molecule is significant as it demonstrates that the non-magnetic toluene in its neutral state becomes paramagnetic in its ionized state, enabling the observation of NMR. In this experiment a hyperfine line at 627 MHz is monitored with $T_d=0.5\mu$sec, nominal $B=$230 G.
We have examined over 100 spectra, all having a sharp line near the expected position of the $^1H$ NMR. A typical NMR spectrum is shown on the whole frequency range in Fig. 3a, and the resonance at 0.96486 MHz is shown in more detail in Figs. 3b, 3c. The gyromagnetic value of $^1H$ 42.577 MHz/T implies that the actual field is 226.6 G.
The scale in Fig. 3a is chosen so a to exhibit the background noise, here just 4\% of the $^1H$ resonance intensity is shown. The signal to noise ratio near the resonance, as shown in Fig. 3a, is $\approx 1000$. 
Within a  total acquisition time of 1 second we recorded  $2\cdot 10^6$ points which allows detection of chemical shifts. This is shown in the very high resolution Fig. 3c that shows splitting of $\sim 10$ppm, as in standard data \cite{toluene}; in this figure the data has been smoothed to avoid the inherent discreteness of 1 Hz. Fig. 3d, at a higher magnetic field, shows splittings of $\sim 100$ppm, possibly chemical shifts which can be larger in radical molecules \cite{canters}, such as ionized toluene.

\begin{figure}[t]
\includegraphics*[height=.33\columnwidth]{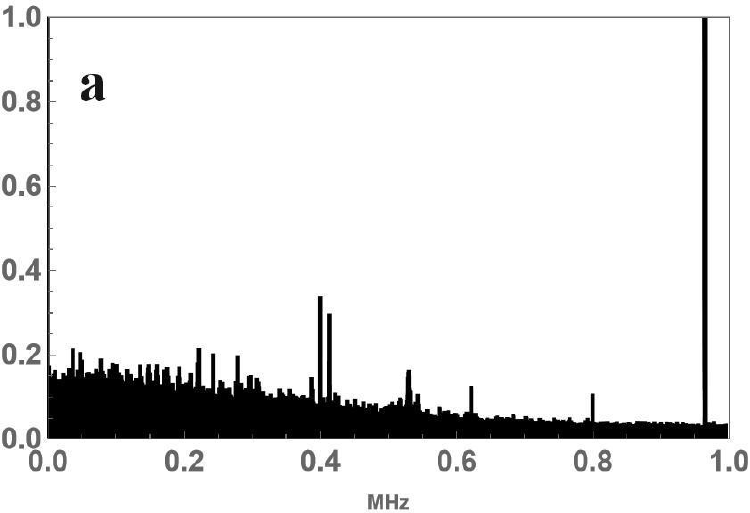}
\includegraphics*[height=.33\columnwidth]{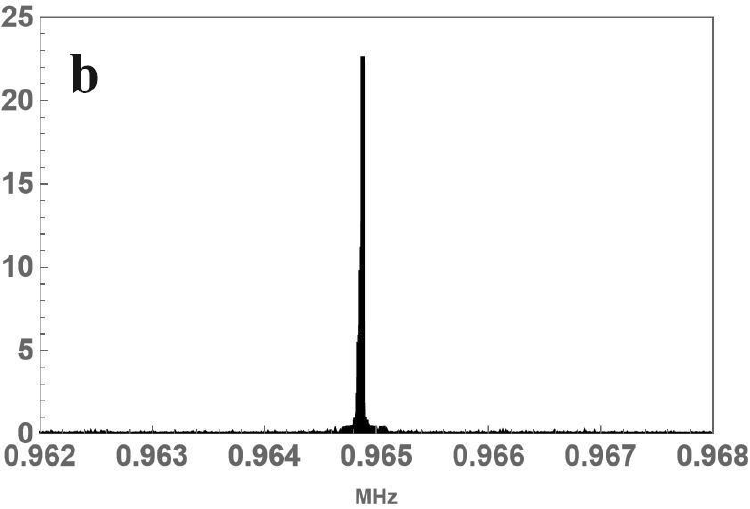}
\includegraphics*[height=.3\columnwidth]{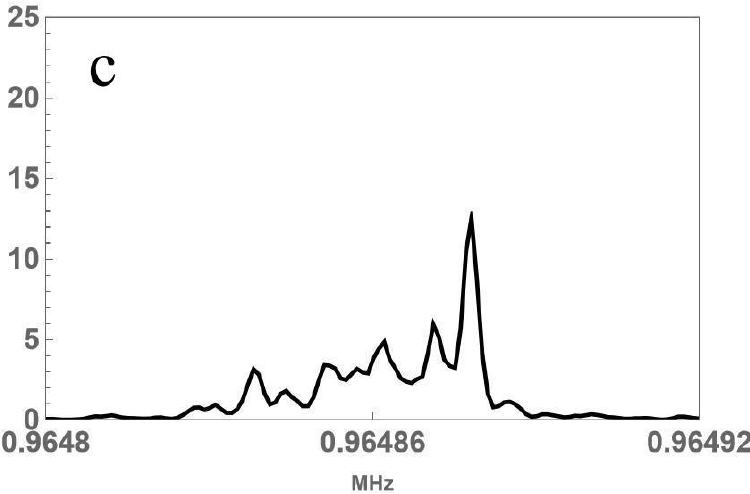}
\includegraphics*[width=.33\columnwidth]{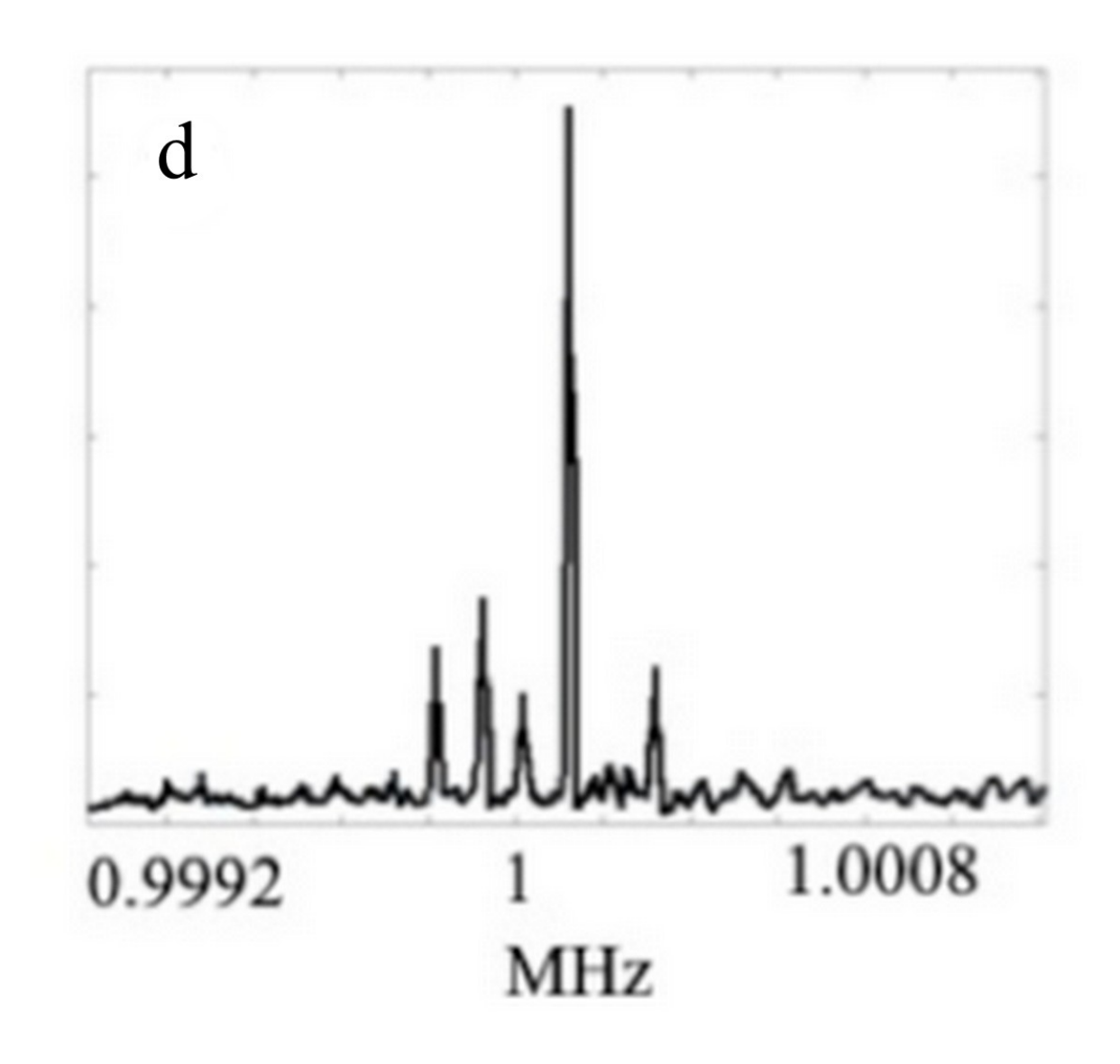}
\caption{{\bf Data on toluene}. (a) Single nucleus NMR of $^1$H in toluene, the signal near 0.97 MHz is shown with just 4\% of its intensity. The scale is chosen so as to exhibit the background noise. (b) Full size of the signal in Fig. 3a. (c) A high resolution spectrum of the signal in (b) showing chemical shifts between aromatic and aliphatic hydrogen peaks. Voltage modulation is between 0.2V and 3.7 V at a frequency of 15.675 MHz, I=0.1 nA, B=230 G, $T_d=0.25\mu$sec. (d) Same parameters except $B=235$ G, demonstrating that the NMR peak indeed shifts with $B$.}
\end{figure}

     We consider next our data on triphenylphosphine molecules in Fig. 4 showing resonaces of $^{13}$C, $^{31}$P and $^{1}$H. Spectra are taken in groups of 50, repated 4 times. Fig. 4a shows the full spectrum with dashed lines indicating the expected resonances at $B=225$ G, using the gyromagnetic factors 10.7084 MHz/T of $^{13}$C, 17.235 MHz/T of $^{31}$P and 42.577 MHz/T of $^{1}$H; we note deviations of up to $\sim$4\%. We also note that the 1.1\% natural abundance of $^{13}$C is compensated by the presence of 18 C atoms in this molecule. The triplet of the $^{31}$P with splitting of 20 Hz is consistent with chemical shifts in similar molecules \cite{rothwell}.

\begin{figure}[t]
\includegraphics*[height=.35\columnwidth]{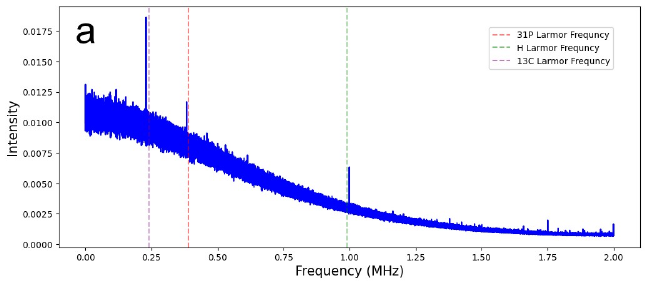}
\includegraphics*[height=.35\columnwidth]{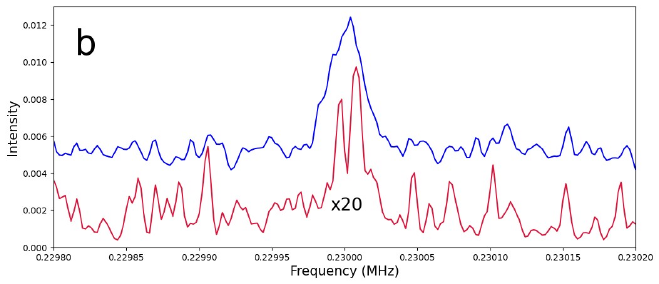}
\includegraphics*[height=.35\columnwidth]{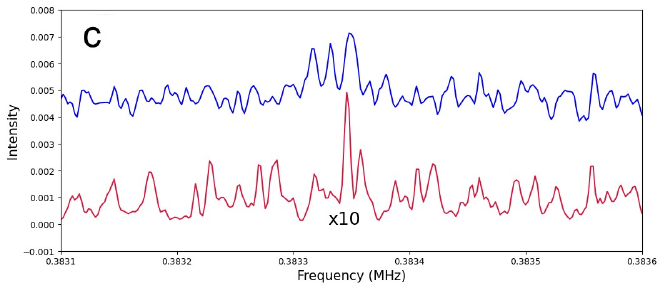}
\includegraphics*[height=.35\columnwidth]{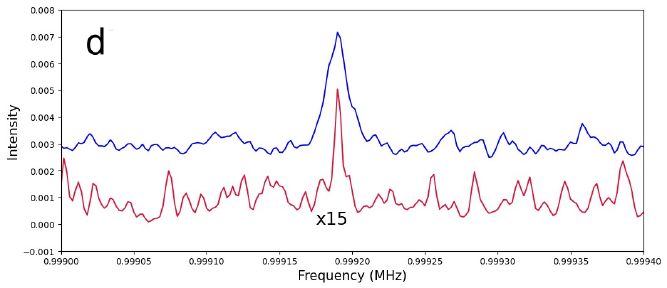}
\caption{{\bf Data on triphenylphosphine}. NMR data using the fast scope method. Monitored ESR frequency is 734 MHz, voltage is modulated between 0.2V and 3.8V at 16.5 MHz, tunneling current is I=0.5 nA, B=225 G. Data is averaged over 50 consecutive spectra. High resolution data is shown for the $^{13}$C line (b), the $^{31}$P line (c) and the $^1$H line (d). The lower lines in red (4a, 4b, 4c) show individual spectra, enhanced as indicated.}
\end{figure}

\begin{figure}[tbh]
\includegraphics*[height=.25\columnwidth]{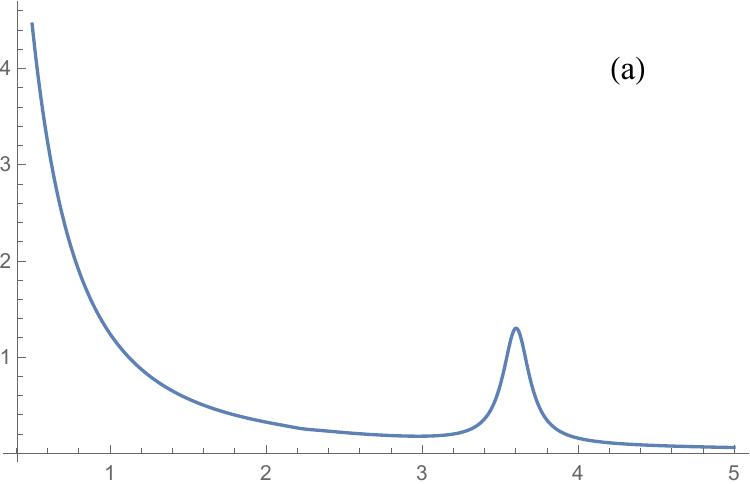}
\includegraphics*[height=.25\columnwidth]{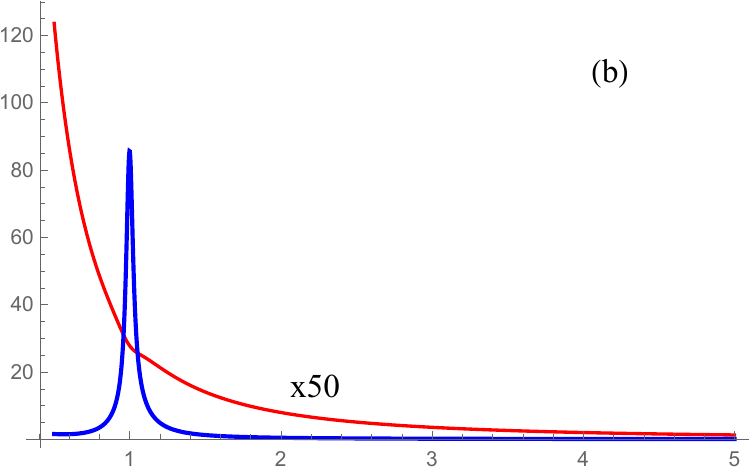}
\includegraphics*[height=.25\columnwidth]{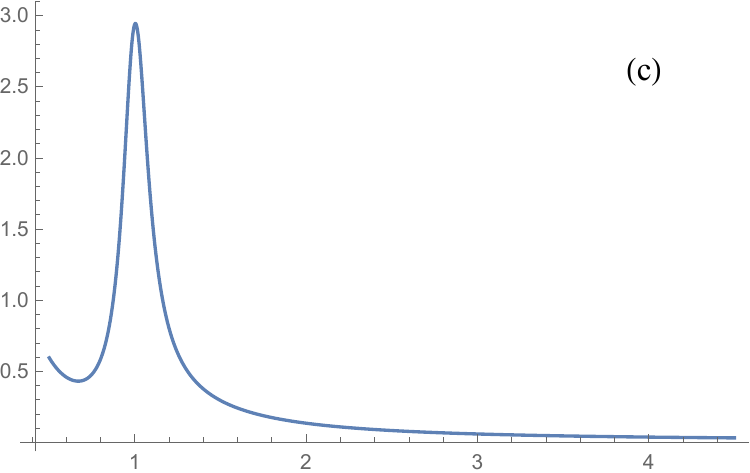}
\includegraphics*[height=.25\columnwidth]{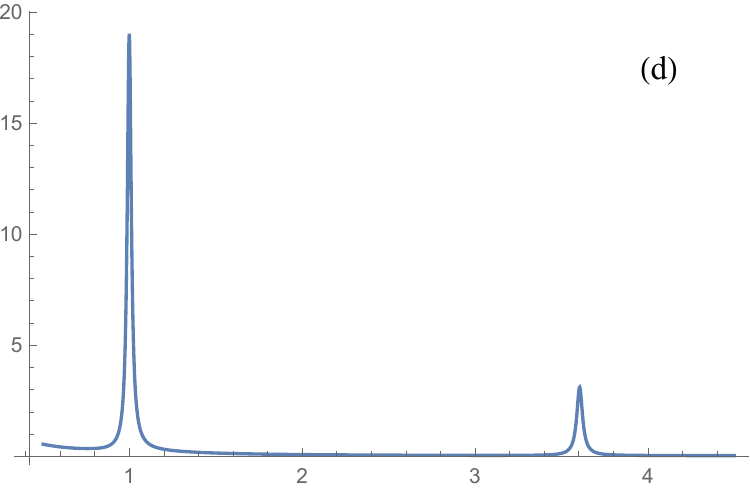}
\caption{{\bf Theory cases}. Theory results showing nuclear correlations with NMR observed at a nuclear frequency $\nu_n=1$ MHz, electron frequencies are 600 MHz (hyperfine coupled) and 650 MHz (spectator). (a) Low $\gamma=0.01$ and low $a=d=1$, $\gamma_1=.1$, (b) case of two nuclei with low $\gamma=0.1$:  high $a=d=10$ (red line, data enhanced by a factor 50) coexisting with a low $a_2=d_2=0.5$ (blue thick line), $\gamma_1=20$. (c) High $\gamma=10,000$ and high $a=d=10$, $\gamma_1=.01$, (d) High $\gamma=10,000$ and low $a=d=1$ , $\gamma_1=.01$.}
\end{figure}
\vspace{2cm}

{\bf Theoretical Model}

We have developed a model to describe the ionization process. The model assumes the presence of a second molecule in the system that collects the electron that is being ionized. A second radical is actually essential for observing ESR-STM \cite{bh}. The basis set has then the state $|11\rangle$ where the two electron spins are on separate molecules while one of the spins has a hyperfine coupling with a nuclear spin $\half$. The $|11\rangle$ state has therefore 8 spin components. In addition, we have the state where both spins are on the same molecule, labeled as $|20\rangle$. The electron spins are then in a singlet state and the only spin components are the two nuclear ones, hence a total of 10 states in the Hilbert space. In the $|11\rangle$ state the nuclear spin dynamics is dominated by the hyperfine, hence free nuclear rotation is allowed only in the $|20\rangle$ state.
We have also considered a case where two nuclear spins are hyperfine coupled, leading to a Hilbert space of 20 states. We build a master equation to describe dissipative transitions of strength $\gamma$ between the $|11\rangle$ and $|20\rangle$ subspaces, as well as dissipative transitions $\gamma_1$ between the two spin states of either electron spin. We find that an off-diagonal hyperfine element is essential for observing the NMR signature, i.e. $d\sigma_z\tau_y$. In the SM \cite{SM} we demonstrate that a rotated molecule has in general such a term. We have extended the model to allow distinct relaxation rates from $|11\rangle$ to $|20\rangle$ and back, to account for a voltage modulation. We find the the modulation has a weak effect on the NMR features, hence the main ingredient are the states $|11\rangle$ and $|20\rangle$ that coexist even in the steady state.

We find two situations when an NMR signal is observed, as shown in Fig. 5, for details see SM \cite{SM}. (i) Case of small $\gamma$: when $a=d=1$ and $\gamma_1=0.1$ there is a strong signal at a shifted position  $\tilde\nu_n=\sqrt{(\nu_n+2a)^2+4d^2}$, Fig. 5a,  which we consider as an extreme case of a chemical shift. It corresponds to a fully polarized $\sigma_z\rightarrow +1$; if $\gamma_1$ increases eventually $\langle\sigma_z\rangle\rightarrow 0$ and this peak approaches $\nu_n$. At the expected $a=d\approx 10$ for toluene the signal shows a weak dip. However, when 2 nuclei are present with different hyperfine couplings, $a=d=10$ and $a_2=d_2=0.5$, e.g. two inequivalent $^1$H nuclei in toluene \cite{wertz}, we observe a strong signal of the weaker coupled nucleus, while a very weak dip of the strongly coupled nucleus persists, Fig. 5b.
 (ii) Case of large $\gamma$ (though still small on the voltage scale): here the singlet state of $|11\rangle$ decays fast and ESR is then possible only between the triplet states of $|11\rangle$. The ESR is then at shifted position while the NMR signal is strong for $a=d=10$ as well as for $a=d=1$, the latter case shows an additional peak, a chemical shift to the same $\tilde\nu_n$ due to the relatively small hyperfine coupling, Figs. 5c,5d. 
 
 We propose that the case with two nuclei, Fig. 5b, is the most likely scenario to account for the experimental data, as it is the only case that allows a relatively large ESR linewidth $\gamma_1$, consistent with the ESR spectrum (Fig. S2). The NMR linewidth $\approx a_2^2/\gamma_1$ (SM \cite{SM} section IIIE) can be identified from the experimental intensity and linewidth.
\\ 

{\bf Conclusions} 

In conclusion, our work demonstrates the successful observation of single spin NMR, a proof of concept which is detected experimentally and accounted for theoretically.  We note that a clear single spin NMR spectrum can be achieved within 1sec, close to the time that a (slow) scan of STM takes to record 1 pixel. This opens the road for observing an atomically resolved STM image concurrently with identifying each nucleus

\acknowledgements
This work was funded by the attract grant “NMR(1)”. Additional funding was provided by the ISF – collaboration with China “ESR-STM study of individual  radical ions at the single molecule level“, the DFG project “Magnetism of Vacancies and edge states in graphene probed by electron spin resonance and scanning tunneling spectroscopy“ and by the EIC grant 101099676 "Single Molecule Nuclear Magnetic Resonance Microscopy for Complex Spin Systems".


\begin{thebibliography}{10}
\bibitem{mamin}  H. J. Mamin, M. Poggio, C. L. Degen, andD. Rugar, Nuclear magnetic
resonance imaging with 90-nm resolution. Nat. Nanotechnol. {\bf 2}, 301–306(2007).
\bibitem{degen} C. L.  Degen, M. Poggio, H. J. Mamin, C. T. Rettner and D. Rugar, Nanoscale
magnetic resonance imaging. Proc. Natl Acad. Sci. {\bf 106}, 1313–1317 (2009).
\bibitem{muller}   C. Muller et al. Nat. Commun. Nuclear magnetic
      resonance spectroscopy with single spin sensitivity. Nat. Comm {\bf 5}, 4703 (2014) .
\bibitem{pfender} M. Pfender et al. High-resolution spectroscopy of single nuclear spins via sequential weak measurements. Nat. Commun. {\bf 10}, 594 (2019). 
\bibitem{manassen1} For a review see A. V. Balatsky, M. Nishijima and Y. Manassen, ESR-STM. Adv. Phys. {\bf 61}, 117 (2012).
 \bibitem{manassen2} Y. Manassen, M. Averbukh and M. Morgenstern, Analyzing multiple encounter as a possible origin of ESR signals in STM on Si(111) featuring C and O defects. Surf. Sci. {\bf 623}, 47 (2014).
 \bibitem{manassen3} Y. Manassen, M. Averbukh, M. Jbara, N. Siebenhofer, A. Shnirman and B. Horovitz, Fingerprints of single nuclear spin energy levels using STM-Endor. J. Magn. Reson. {\bf 289}, 107 (2018).
 \bibitem{baumann} S. Baumann, W. Paul, T. Choi, C. P. Lutz, A. Ardavan and A. Heinrich, Electron Paramagnetic Resonance of individual atoms on a surface. Science {\bf 350}, 417 (2015).
 \bibitem{willke} P. Willke et al. Probing quantum coherence in single-atom
electron spin resonance, Sci. Adv. {\bf 4}, eaaq1543 (2018).
\bibitem{SM} see Supplementary Material.
 \bibitem{tolman}  J. R. Tolman and K. Ruan, NMR Residual Dipolar
       Couplings as Probes of Biomolecular Dynamics. Chem. Rev. {\bf 106}, 1720-1736 (2006).
 \bibitem{hazan} Z. Hazan, M. Averbukh and Y. Manassen, ESR-STM on Diamagnetic Molecule: C60 on graphene. J.  Mag. Res. {\bf 348}, 107377 (2023).
 \bibitem{dimer} .Y. Manassen, M. Jbara, M. Averbukh, Z. Hazan, C. Henkel and B. Horovitz, Phys. Rev. B {\bf 105}, 235438 (2022).
 \bibitem{witanowski} M. Witanowski and G. A. Webb, Nitrogen NMR. Plenum Press (London and NY, 1973)
 \bibitem{toluene} url: https://www.google.com/search?client=firefox-b-d\&q=Toluene+NMR\#imgrc=S0bx7852rKbZeM
 \bibitem{canters} G. W. Canters 
and E. de Boer, Mol. Phys. {\bf 13} 395 (1967).
\bibitem{rothwell} W. P. Rothwell, W. X. Shen and J. H. Lunsford, J. Am. Chem. Soc. {\bf 106}, 2452 (1984).
 \bibitem{bh} B. Horovitz and A. Golub, Double quantum dot scenario for
spin resonance in current noise. Phys. Rev. B {\bf 99}, 241407(R) (2019).
\bibitem{wertz} J. E. Wertz and J. R. Bolton,  Electron Spin Resonance Elementary Theory and Practical Applications. Chapman \& Hall (1986)
\end{thebibliography}
\end{document}


\title{NMR of a single nuclear spin detected by a scanning tunneling microscope \\ \vspace{.0cm}
Supplementary Material}
\author{Yishay Manassen$^1$, Michael Averbukh$^1$, Zion Hazan$^1$, Yahel Tzuriel$^1$, Pino Boscolo$^2$, Alexsander Shnirman$^3$ and Baruch Horovitz$^{1}$}
\affiliation{$^1$Department of Physics, Ben Gurion University of the Negev, Beer Sheva 84105, Israel\\
 $^2$Gruppo Techniche Avanzate, Via Vergerio 1, 34138 Trieste, Italy\\
 $^3$Institut für Theorie der Kondensierten Materie, Karlsruhe Institute of Technology, D-76131 Karlsruhe, Germany}
 \maketitle

\makeatletter
\renewcommand{\thefigure}{S\@arabic\c@figure}
\makeatother

\section{Materials and Methods}

  Our experiments were carried out with a Demuth type STM operated at room temperature in ultrahigh vacuum (UHV) (base pressure $\leq 1.5\cdot 10^{-10}$ Torr). STM images were acquired with chemically etched tungsten tip (W). The molecules were deposited on gold films of thickness 100nm on Mica. The deposition was done in two ways: TEMPO was dissolved in toluene and drop casted on the surface at a concentration of 0.041 g/25 ml, corresponding to one monolayer. Toluene and triphenylphosphine were evaporated for few minutes with a leak valve maintaining base pressure of $10^{-8}$ Torr in the chamber. Molecular resolution was achieved, and it was easy to identify single molecules on the surface (Fig. 1). The next step was to study their magnetic resonance signature.

\begin{figure}[tbh]
\includegraphics*[height=.25\columnwidth]{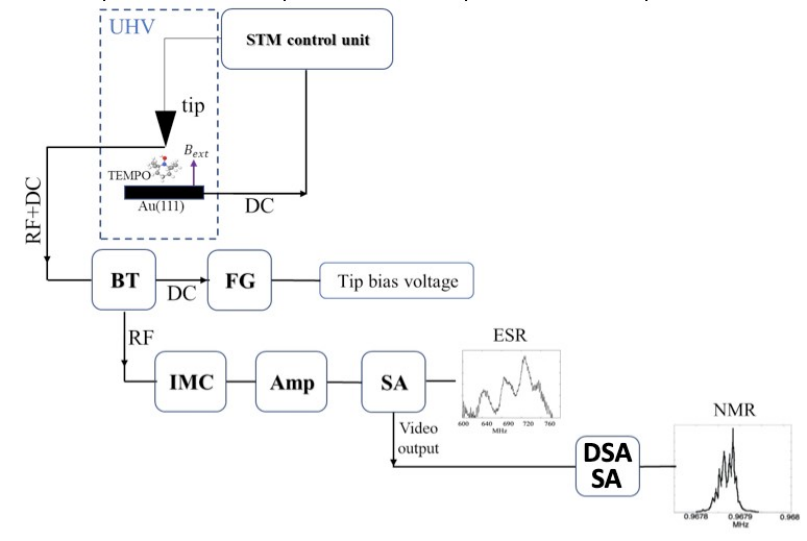}
\caption{{\bf Setup.} NMR-STM acquistion setup.}
\label{setup}
\end{figure}

  The electronic setup for our ESR and NMR data is shown in Fig. \ref{setup}. DC tunneling current from the sample was connected to the STM control unit for the STM image acquisition. RF and DC tunneling currents from the tip were split with a bias-tee (BT) where the DC part (frequency $f<30$MHz) was connected to a frequency generator (FG) that modulated the STM tip bias voltage ($0.1<V_{bias}<4$V) with the desired modulation frequency ($0.25<f<30$MHz). The RF part was connected to an impedance matching circuit (IMC) and to an amplifier (Amp). A spectrum analyzer (SA) recorded the RF intensity as a function of frequency (span: $200<f<800$MHz) at constant magnetic field, $B_{ext}\approx 230$G, and the output was the ESR-STM spectrum. Then, the intensity of one of the hyperfine peaks (with a bandwidth of 3MHz) was digitally recorded and finally its power spectrum calculated (DSA) as a function of frequency, the output was the NMR-STM spectrum.

We have also studied a different two stage experiment: one spectrum analyzer records a chosen hyperfine line intensity as function of time, generating a video output. A second spectrum analyzer is directly connected to find the power spectrum of that intensity at the low NMR frequencies (i.e. SA at the bottom of Fig. \ref{setup} instead of DSA). The advantage of this method is that a high resolution NMR spectrum is seen immediately without time consuming calculations of the results.  The disadvantage is that its spectral resolution is relatively low.
This can be improved by either a new spectrum analyzer with a shorter dwell time, or by using a fast scope with a short dwell time.

\begin{figure}[t]
\includegraphics*[height=.25\columnwidth]{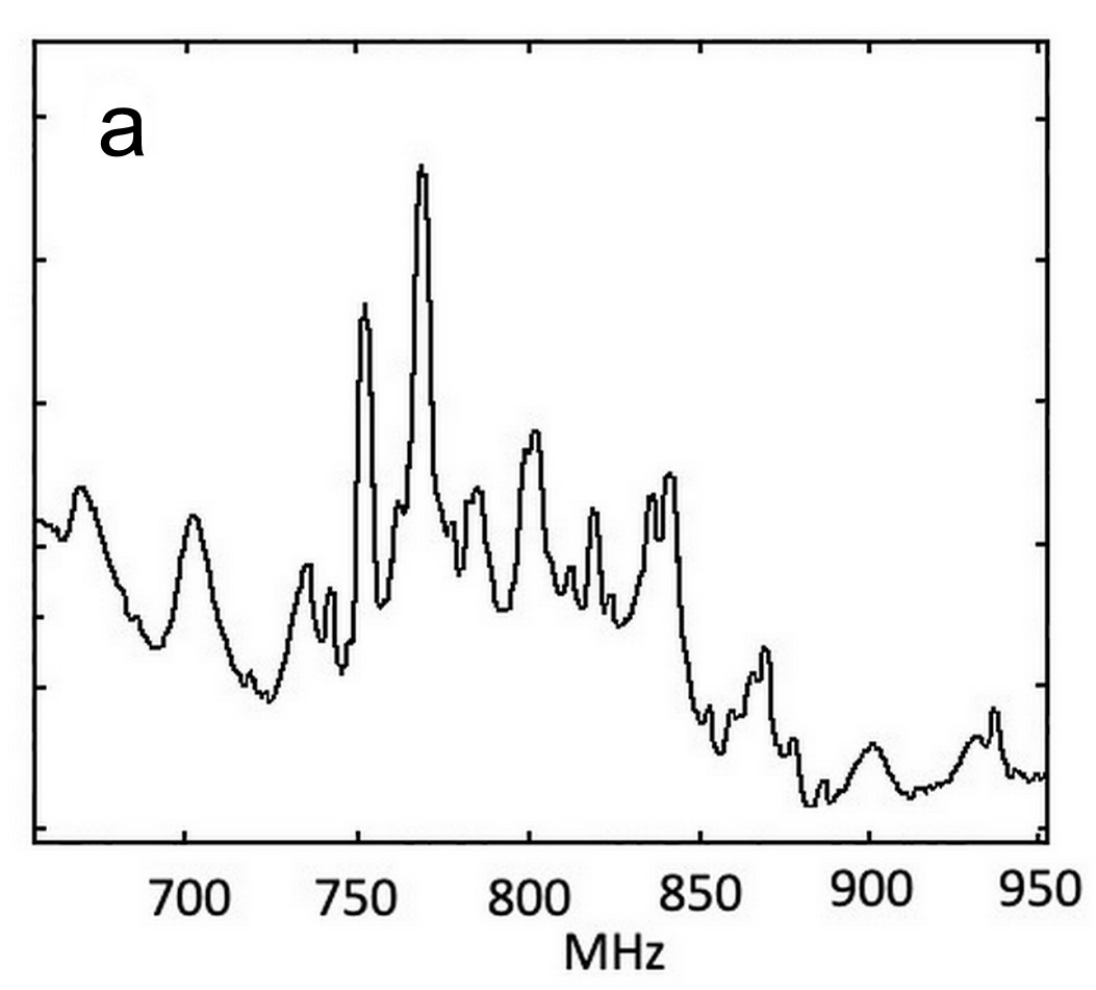}
\includegraphics*[height=.25\columnwidth]{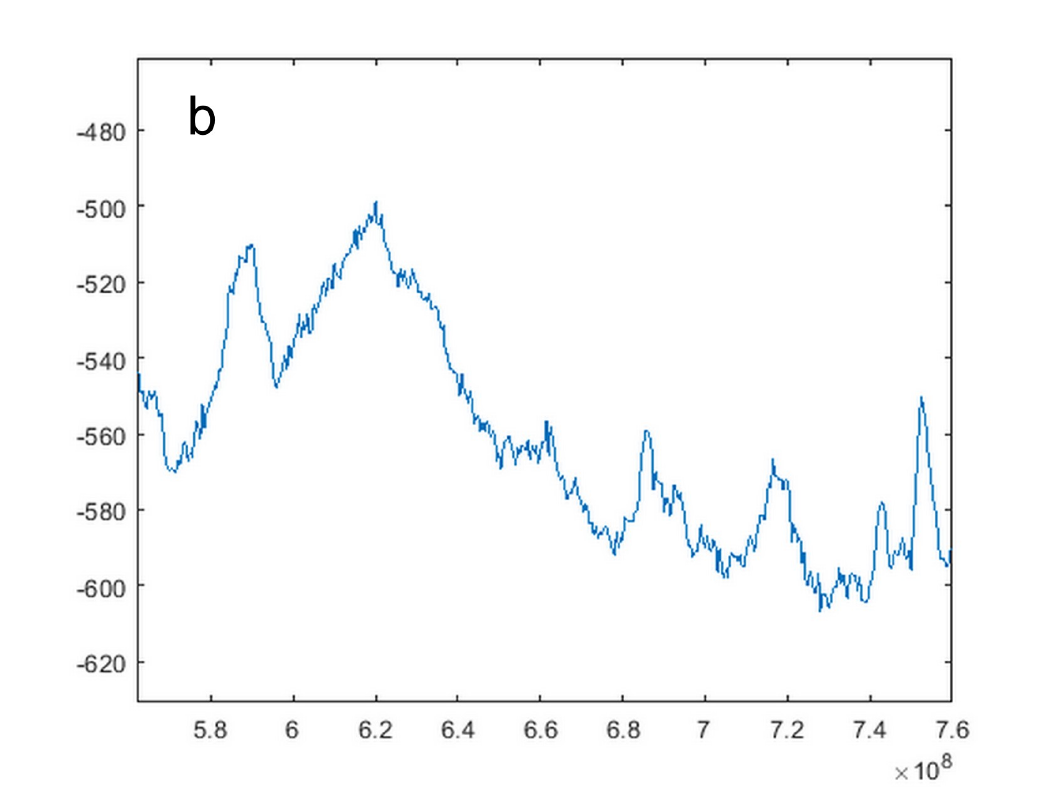}
\includegraphics*[height=.25\columnwidth]{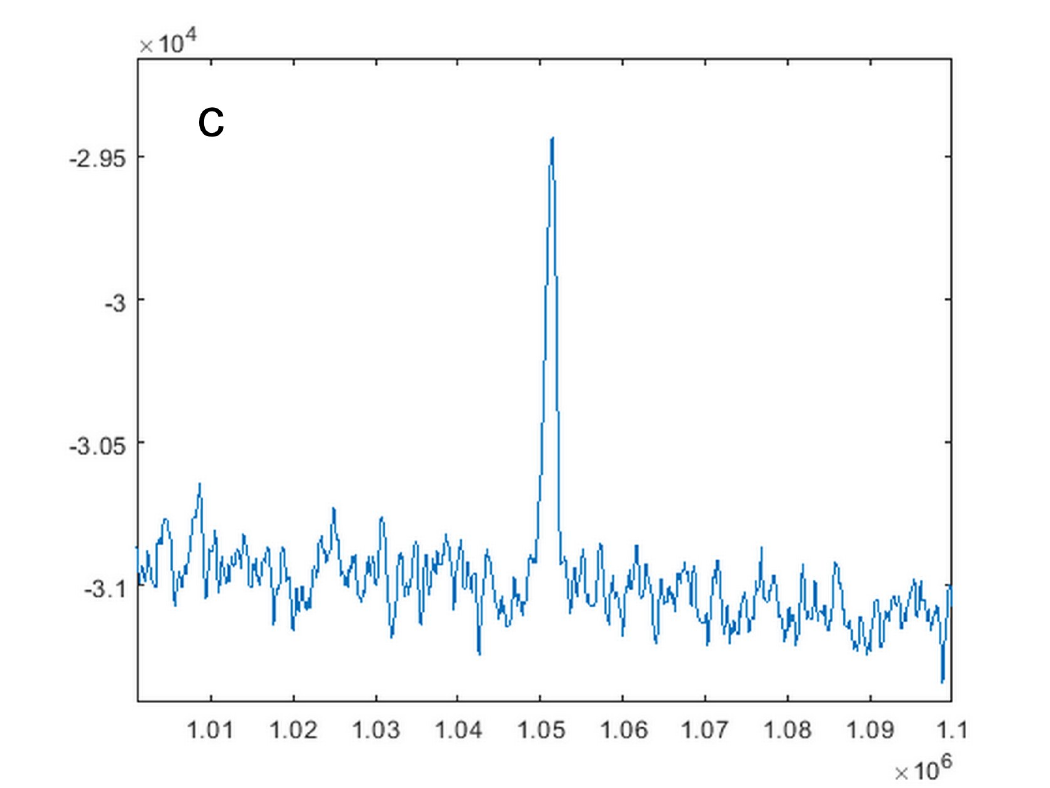}
\caption{{\bf Data on toluene.} (a) ESR data with a constant voltage of 2V, I=0.1 nA, corresponding to B=286 G. (b)
ESR data with bias voltage modulation of 0.2V and 3.8V at a frequency of 15MHz, I=1nA, B=235G. (c) NMR spectrum using ESR data at 660MHz is chosen and analyzed by a second spectrum analyser.}
\label{supFig2}
 \end{figure}

The results of this procedure are shown at Fig. \ref{supFig2}. The sample as in the main text, is a monolayer of toluene on a gold surface. ESR spectra are shown in Fig \ref{supFig2}a at constant voltage and in \ref{supFig2}b with voltage modulation, showing that it is well defined even in letter case. We note that the distance between the peaks is 17MHz, as expected for toluene radical anion \cite{tuttle}.
With B=235G the NMR signal of $^1$H is expected at 1.001MHz, indeed close to the peak shown in Fig. \ref{supFig2}b. We note that the intensity of the NMR peaks are smaller in this method.

 We have carried out yet another type of experiment on TEMPO, where the low frequency hyperfine peak intensity in the ESR is analyzed by a lock in amplifier in which the reference frequency was swept from 0 to 150 kHz. A significant signal was detected quite often at 70 kHz, consistent with the Larmor frequency of the $^{14}$N nucleus at 230 G (Fig. S3a); an additional signal was observed at half of this frequency. In this experiment a single sweep has taken 90 sec, a long time that caused a linewidth much broader than those with the fast scope method as shown in the main text and in Fig. S3b.  In the latter figure we have taken 5 consecutive spectra over a given molecule, showing line positions that vary by $\sim 0.1\%$, possibly due to a slight drift of either the magnetic field or the tip position.

     \begin{figure}[b]
 	\includegraphics*[height=.25\columnwidth]{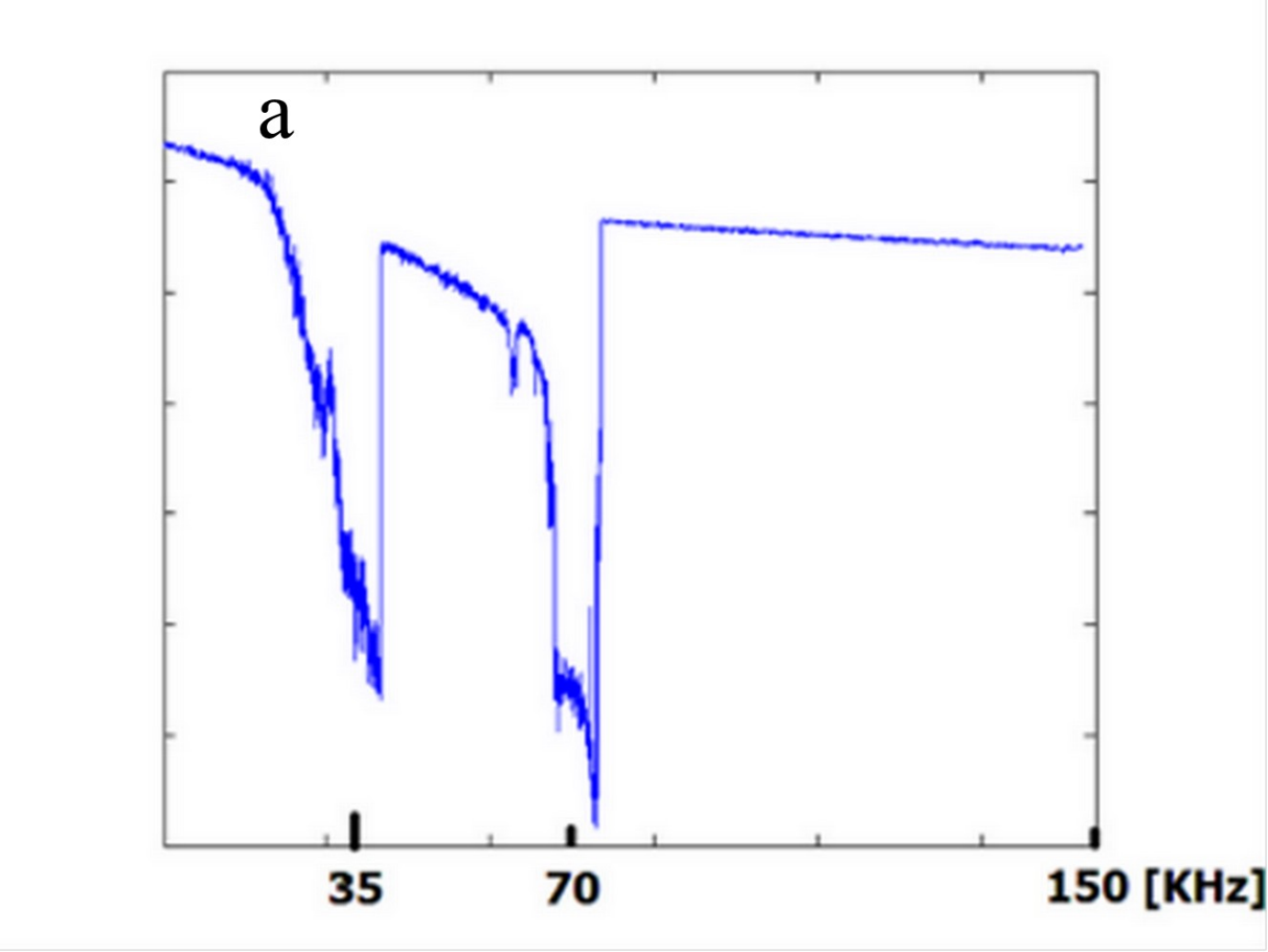}
 	\includegraphics*[height=.25\columnwidth]{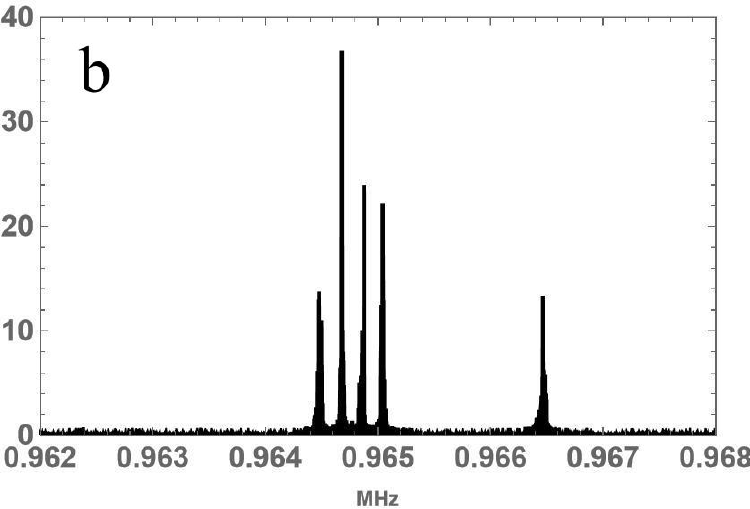}
 	\caption{{\bf Data on Tempo.} (a)    NMR data using a lock-in method, monitoring the low frequency hyperfine peak. Voltage is modulated between 0.2V to 3.7V at a frequency of 250 kHz, I=0.1 nA, B=230 G.
 	(b) Five consecutive measurements of the 1H resonance with the fast scope method. These lines include the one shown in Fig. 3b, parameters are the same as in Fig. 3.}
 	\label{supFig3}
 \end{figure}

 We next summarize the parameters of the experiment, that will be incorporated in the following theory sections. These parameters also appear in the captions of figures in the main text. The parameters are: Magnetic field $B$, chosen in the $z$ direction; the STM bias voltage chosen as sample positive (i.e. the tip is at negative voltage with respect to the substrate), if the voltage is modulated with period $\bar T$ it has high and low voltage values, each during time $T_1=T_2=\bar T/2$; the tunneling current $I$; a chosen hyperfine frequency $\nu_{hyper}$ that is detected by a spectrum analyzer with a bandwidth of 3MHz; the intensity of this hyperfine signal is detected during a dwell time $T_d$. We note that $T_d=0.5\mu$sec for the data in Fig. 3 and $T_d=0.25\mu$sec for Figs. 2,4, for the lock-in method (Fig. 2c) $T_d$ is in some sense its response time $\approx 100\mu$sec, though this implies a too large $\nu_nT_d$. The second spectrum analyzer method (Fig. \ref{supFig2}c) acquires time dependent data continuously so $T_d$ is not well defined, indeed the data in Fig. \ref{supFig2}c is rather noisy. $T_d$ should be in the range $\nu_n\ll 1/T_d\ll \nu_{hyper}$ so that $\nu_{hyper}$ can be detected accurately (many ESR oscillations within $T_d$) while the nuclear polarization is almost constant (almost no NMR oscillation within $T_d$).

\section{Rotated hyperfine}
Suppose that the hyperfine tensor has principal axes $z'y'x'$ at some orientation relative to the magnetic field, chosen in the $z$ direction. The diagonal hyperfine elements are $c',b',a'$. We wish to rotate this tensor and find the dominant hyperfine splitting for electron Larmor frequency $\nu\gg a',b',c'$.  Assume for simplicity that it is sufficient to rotate around the $x=x'$ axis with an angle $\theta$, this is the case if $c'$ is relatively small or if two elements are equal $b'=c'$. The hyperfine coupling with the rotation matrix $R(\theta)$ becomes, where the electron spin is $\bf S$, the nuclear spin is $\bf I$ and primes denote rotated quantities,
\beq{04}
S'\cdot A'\cdot I'=&&(S_z,S_y,S_x)\left(\begin{array}{ccc} \cos\theta &\sin\theta & 0 \\ -\sin\theta & \cos\theta & 0 \\
0 & 0& 1 \end{array}\right)\left(\begin{array}{ccc} a' & 0 &0 \\0 & b' & 0 \\ 0 & 0& c'\end{array}\right)
\left(\begin{array}{ccc} \cos\theta &-\sin\theta & 0 \\ \sin\theta & \cos\theta & 0 \\
0 & 0& 1 \end{array}\right)\left(\begin{array}{c} I_z \\ I_y \\ I_x \end{array}\right)\nonumber\\&&
=(S_z,S_y,S_x)\left(\begin{array}{ccc} a'\cos^2\theta+b'\sin^2\theta & \sin\theta\cos\theta(b'-a') & 0 \\
\sin\theta\cos\theta(b'-a') & a'\cos^2\theta+b'\sin^2\theta & 0 \\ 0 & 0 & c' \end{array}\right)
\left(\begin{array}{c} I_z \\ I_y \\ I_x \end{array}\right)
\eeq
where the rotated spin vectors ${\bf S}={\bf S}'\cdot R(-\theta),\, {\bf I}=R(\theta)\cdot {\bf I}'$. For a general rotation all elements of the tensor become finite. In presence of a Larmor frequency $\nu$ in the Hamiltonian, i.e. $ \nu S_z$, the $S_x,\,S_y$ terms are perturbative of order (hyperfine)$^2/\nu$, hence the dominant hyperfine term is
\beq{05}
&&{\cal H}_{hyp}=S_z\,(aI_z+dI_y)=\sqrt{a^2+d^2} S_z\, \tilde I, \qquad
\tilde I=\frac{a}{\sqrt{a^2+d^2}}I_z+\frac{d}{\sqrt{a^2+d^2}}I_y\\&&
a=a'\cos^2\theta+b'\sin^2\theta,\qquad d=\sin\theta\cos\theta(b'-a'),\qquad \Rightarrow a^2+d^2=a'^2\cos^2\theta+b'^2\sin^2\theta\nonumber
\eeq
The eigenvalues $\pm\sqrt{a'^2\cos^2\theta+b'^2\sin^2\theta}$ are in agreement with Eq. 1.66 (or  Eq. 3.44) of Ref. \onlinecite{abragam}(g factor is assumed isotropic).

\section{Model and simulations}

\subsection{The model}

There are two essential ingredients in our theoretical model: (i) In addition to the tested molecule the system has an additional spectator molecule that can absorb an ionized electron from the tested one. This leads to the states $|11\rangle$ and $|20\rangle$ as defined in the main text. We assume that these two states coexist, the coexistence may be affected by a voltage modulations. (ii) The hyperfine tensor has a off diagonal element, a common possibility as shown in the previous section.

We note that the tested molecule can be either a radical or a non-radical at equilibrium. In the radical case, e.g. TEMPO, in the neutral state both molecules carry a spin forming a $|11\rangle$ state, while in the ionized state the TEMPO electron, i.e. the spin carrying one, is transferred to the spectator molecule forming there a singlet state, i.e. the $|20\rangle$ state. When the tested molecule is a non-radical, e.g. toluene, the roles are reversed -- in the neutral state both molecules are non-radicals forming a $|20\rangle$ state, while in the ionized state one toluene electron from a singlet state is transferred to the spectator molecule so that both molecules become radicals, i.e. the $|11\rangle$ state. Depending on the STM voltage, it is also possible that one spectator electron is transferred to toluene, also forming a $|11\rangle$ state. Our model treats all cases on equal footing.
 Evidently, the measured ESR signal comes from the $|11\rangle$ state while the derived NMR come from the coexisting $|20\rangle$ state. The situation is somewhat similar to that of two quantum dots \cite{petta} that by changing gate voltages can transfer charge between their $|20\rangle$ and $|11\rangle$ states.

The $|20\rangle$ state has two states due to the nuclear spin, while $|11\rangle$ has eight states (two electrons and one nuclear having each 2 states). Hence the Hilbert space has 10 states.
The Hamiltonian in the two subspaces, respectively, and the total Hamiltonian are
\beq{06}
&&{\cal H}_{11}=\half\nu_1\sigma_z\otimes\mathbbm{1}\otimes\mathbbm{1} +\half\nu_2 \mathbbm{1}\otimes s_z \otimes \mathbbm{1} +\half\nu_n\mathbbm{1}\otimes\mathbbm{1}\otimes\tau_z+a\sigma_z\otimes\mathbbm{1}
\otimes\tau_z+b\sigma_x\otimes\mathbbm{1}\otimes\tau_x +d\sigma_z\otimes\mathbbm{1}\otimes\tau_y \nonumber\\&&
{\cal H}_{20}=\half\nu_n\tau_z\nonumber\\&&
{\cal H}={\cal H}_{11}|11\rangle\langle 11|+{\cal H}_{20}|20\rangle\langle 20|
\eeq
where ${\bm \sigma},\,{\bf s},\,{\bm\tau}$ are Pauli matrices representing the two electron spins and the nuclear spin, respectively and $\nu_1,\,\nu_2$ are the electron resonance frequencies of the two radicals, respectively (in the absence of hyperfine couplings). An additional direct tunneling element between $|11\rangle$ and $|20\rangle$ is possible, as well as a chemical potential shift between these subspaces; we find that both terms have a minor effect on the results below.

 We assume the switching between the $|11\rangle$ and $|20\rangle$ subspaces to be dissipative. We introduce the rate $\gamma$ for the $|20\rangle\rightarrow|11\rangle$ transition and $\gamma_d$ for the reverse process. Labeling the states with $\pm$ for $\sigma_z$, $s_z$ and $\tau_z$ spins, respectively, we order the $|11\rangle$ space as $|+++,++-,+-+,+--,-++,-+-,--+,---\rangle$. The jump operators couples only the singlet component of the $|11\rangle$ state, which is either $|0, 0, 0, -1, 0, 1, 0, 0 \rangle/\sqrt{2}$ or $|0, 0, -1, 0, 1, 0, 0, 0\rangle/\sqrt{2}$ (i.e. two nuclear spins). We then construct a $10\times 10$ operator L that transfers the $|20\rangle$ singlet states into $|11\rangle$, while the transpose $L_d=L^{T}$ ($L^T$ is the transpose of $L$) is transferring the opposite way.
If $\gamma\gg\gamma_d$ the dominant subspace is $|11\rangle$ representing an "ESR" state, while in the opposite case $|20\rangle$ is dominant, representing an "NMR" state. These decay rates depend on voltage, hence when voltage is modulated with a period $\bar T$ then $\gamma>\gamma_d$ during time $T_1$ while $\gamma,\,\gamma_d$ are interchanged during time $T_2$ with $T_1+T_2=\bar T$.

Since only the singlet states are populated by $\gamma,\,\gamma_d$ we need additional relaxations within the $|11\rangle$ states, so as to populate all states. The resonance frequencies $\nu_1$ or $\nu_2$ are much smaller than temperature (or voltage), hence spin flip rates up or down are equal with values $\gamma_1,\,\gamma_2$, for the two radicals, respectively. We define a $10\times 10$ operator $M_1=\sigma_+\otimes\mathbbm{1}\otimes\mathbbm{1}$ and 0 in  the $2\times 2$ space of $|2 0\rangle$, similarly with $s_+\rightarrow M_2$. Hence
 the Lindblad equation during the "ESR" period $T_1$ is ($L,\,M_1,\,M_2$ are real, e.g. $L^\dagger=L^T=L_d$),
\beq{07}
\frac{d\rho}{dt}=&&-i[{\cal H}_1,\rho]+\{\gamma[L\cdot\rho\cdot L^\dagger-L^\dagger\cdot L\cdot\rho]+
\gamma_b[L^\dagger\cdot\rho\cdot L-L\cdot L^\dagger\cdot\rho] \nonumber\\&&+
\gamma_1[M_1\cdot\rho\cdot M_1^\dagger-M_1^\dagger\cdot M_1\cdot\rho +M_1^\dagger\cdot\rho\cdot M_1-M_1\cdot M_1^\dagger\cdot\rho]\nonumber\\&&
+\gamma_2[M_2\cdot\rho\cdot M_2^\dagger-M_2^\dagger\cdot M_2\cdot\rho+M_2^\dagger\cdot\rho\cdot M_2-M_2\cdot M_2^\dagger\cdot\rho]
+h.c.\}
\eeq
During the "NMR" period $T_2$ the roles of $\gamma,\,\gamma_b$ interchange.

\subsection{Master equation}

Consider the density matrix $\rho_{kj}$ as a vector (super-vector) whose elements are lexicographically ordered.
Consider a $10\times 10$ matrix that operates on the density matrix.
To find its corresponding super-operators, which are $100\times 100$, we write an $i,j$ element of regular matrix multiplication as (sums on $k$ or $kl$ are implied, $I_d$ is a $10\times 10$ unit matrix),
\beq{08}
&&A_{ik}\rho_{kj}=(A\otimes I_d)_{ij,kl}\rho_{kl} \qquad \Rightarrow A\cdot\rho= (A\otimes I_d)\cdot \rho\qquad \text{product from left}\nonumber\\
&&\rho_{ik} A_{kj}=(I_d\otimes A^T)_{ij,kl}\rho_{kl}\qquad\Rightarrow \rho\cdot A= (I_d\otimes A^T)\cdot\rho\qquad \text{product from right}\nonumber\\
&&(A\cdot\rho\cdot A^\dagger)_{ij}=A_{ik} A^\dagger_{lj}\rho_{kl}=(A\otimes A^{\dagger T})_{ij,kl}\rho_{kl}
\eeq
The Lindblad operators are then
\beq{09}
LL_1=&&-i{\cal H}_1\otimes I_d+iI_d\otimes{\cal H}_1^T+\half\gamma (2L\otimes L_d^T-L_d\cdot L\otimes I_d-I_d\otimes[L_d\cdot L]^T)\nonumber\\&& +\half\gamma_b (2L_d\otimes L^T-L\cdot L_d\otimes I_d-I_d\otimes[L\cdot L_d]^T)\nonumber\\ LL_2=&&-i{\cal H}_2\otimes I_d+iI_d\otimes{\cal H}_2^T+\half\gamma_b (2L\otimes L_d^T-L_d\cdot L\otimes I_d-I_d\otimes[L_d\cdot L]^T)\nonumber\\&& +\half\gamma (L_d\otimes L^T-L\cdot L_d\otimes I_d-I_d\otimes[L\cdot L_d]^T)\nonumber\\
MM_1=&&\half\gamma_1[2M_1\otimes M_1-M_1^T\cdot M_1\otimes I_d-I_d\otimes (M_1^T\cdot M_1)^T\nonumber\\&&
+2M_1^T\otimes M_1^T-M_1\cdot M_1^T\otimes I_d-I_d\otimes (M_1\cdot M_1^T)^T]
\eeq
and similarly for $MM_2$ with $1\rightarrow 2$. The evolution operators are then
\beq{10}
&&L_1=LL_1+MM_1+MM_2,\qquad \frac{d\rho}{dt}=L_1\rho\Rightarrow\qquad \rho(T_1)=\eexp{L_1T_1}\rho(0)\nonumber\\&&
L_2=LL_2+MM_1+MM_2,\qquad \frac{d\rho}{dt}=L_2\rho\Rightarrow \rho(T_1+T_2)=\eexp{L_2T_2}\eexp{L_1T_1}\rho(0)
\eeq
Hence the evolution operator for one modulation period is $U=\eexp{L_2T_2}\eexp{L_1T_1}$. The experiment has, in some cases, an additional period, the dwell time $T_d$, during which the ESR signal is recorded, eventually the list of these intensities generates a power spectrum that shows the NMR signal. In the present formulation we assume that $T_d=\bar T$. The reasoning is that with a voltage modulation $V(t)=\sum_m a_m\cos(2\pi mt/\bar T)$ ($m$ are integers) the $m$-th component produces a signal at $\nu_n+m/\bar T$, since $1/\bar T\gg \nu_n$ in the actual experiments the detection in the vicinity of $\nu_n$ is insensitive to $m\neq 0$, i.e. the time average of $V(t)$ dominates. We therefore consider the evolution $U$ to be within the period $T_d$, yet, we divide this time to $T_1+T_2$ so as to probe the effect of voltage modulation. We find numerically that cases with $\gamma\neq\gamma_d$ are fairly similar to those with $\gamma=\gamma_d$.
In the simulations below we use $T_1=T_1=.05\mu$sec, we find that the results are insesitive to the values of $T_1,\,T_2$ as long as they are in the window $1/\nu_1\ll T_1+T_2\ll 1/\nu_n$.

Define eigenvectors $U\rho_i=\lambda_i\rho_i$, hence an expansion $\rho(0)=\sum_{i=1}^{100}c_i\rho_i$ yields
\beq{11}
\rho(N(T_1+T_2))=\sum_ic_i\lambda_i^N\rho_i\Rightarrow\qquad \langle\tilde \tau(N)\rangle=\sum_ic_i|\lambda_i|^N\eexp{iN\varphi_i}\tr[\tilde \tau \rho_i]
\eeq
We expect that only $i=1$ has $\lambda_1=1$ and $\tr[\rho_1]=1$, i.e. $\rho_1$ is the steady state while for $i>1$
$|\lambda_i|<1$ and $\tr[\rho_i]=0$ (otherwise the steady state is not unique).
Thus we need an eigenvector $\rho_i$ for a particular $i\neq 1$ such that (i) $\lambda_i$ has a phase, equal to $\nu_n\cdot(T_1+T_2)$, (ii) its amplitude $|\lambda_i|$ is very close to 1 so that its linewidth is narrow, and (iii) has a reasonable amplitude $\tr[\tilde \tau \rho_i]$ for NMR observation, $\tilde \tau$ are Pauli matrices that correspond to a nuclear spin $\half$ with operators ${\bf I}$, Eq. \eqref{e05}. This tests the nuclear projection on $\rho_i$.

For the numerical program (using Mathematica), it is more efficient to find the eigenvectors $v^{(1)}_i$ and eigenvalues $e^{(1)}_i$ of $L_1$ and then build a matrix $E_1$ of size $100\times 100$ whose columns are $e^{(1)}_i$. In terms of a diagonal matrix $D_1$ whose elements are $\eexp{e^{(1)}_i T_1}$ the evolution during $T_1$ is
\beq{12}
U_1=E_1D_1(E_1)^{-1}
\eeq
and similarly with $1\rightarrow 2$. To prove this, drop the index 1 or 2 for convenience, i.e. $L\cdot v_i=e_iv_i$, $U=\eexp{LT}$ and define $E_{ij}=(v_j)_i$ and $D_{jk}=\delta_{jk}\eexp{e_k T}$ and check $i,k$ elements of both sides in $U\cdot E=E\cdot D$:
\beq{13}
&&(E\cdot D)_{ik}=\sum_j(v_j)_i\delta_{jk}\eexp{e_kT}=(v_k)_i\eexp{e_kT}\\&&
(U\cdot E)_{ik}=\sum_{n=0}^\infty\sum_j\left(\frac{(LT)^n}{n!}\right)_{ij}(v_k)_j=\sum_{n=0}^\infty \left(\frac{(LT)^n}{n!}\cdot v_k\right)_i=\sum_{n=0}^\infty \left(\frac{(e_kT)^n}{n!}v_k\right)_i=\eexp{e_kT}\left(v_k\right)_i\nonumber
\eeq
Note that the eigenvectors are not orthogonal ($L_1$ is not hermitian), we need, however, that $E_1$ is invertable, otherwise a subset of $\rho_i$ can span the whole space.

\subsection{ESR}

Consider first the electron correlation $C_i(\omega)$, assuming that the initial state is some $\rho_i$ so that after $N$ periods $\rho_i(N(T_1+T_2))=\lambda_i^N\rho_i$. For $i>1$ this decays with $N$, yet it indicates effects in the correlation between different $N$'s; for the steady state we need $C_1(\omega)$.
Using the identities \eqref{e08} for transforming into supermatrices and using $C_{-+}(-t)=\langle \sigma_-(-t)\sigma_+(0)\rangle=
\langle\sigma_-(t)\sigma_+(0)\rangle^*=C_{-+}^*(t)$, we have
\beq{14}
&&C_i(\omega)=\int_0^\infty\langle\sigma_-(t)\sigma_+(0)\rangle\eexp{i\omega t}dt+
c.c.
=\lambda_i^N\int_0^\infty\tr[\eexp{i{\cal H}t}\sigma_-\eexp{-i{\cal H}t}\sigma_+\rho_i]\eexp{i\omega t}+c.c.
\nonumber\\&&
=\lambda_i^N\int_0^\infty\tr[(\sigma_-\otimes I_d)\eexp{L_1t}(\sigma_+\otimes I_d)\rho_i]\eexp{i\omega t}+
c.c.
=\lambda_i^N \bar C_i(\omega)+c.c.\nonumber\\&&
\bar C_i(\omega)\equiv\tr[(\sigma_-\otimes I_d)\frac{1}{-i\omega-L_1}(\sigma_+\otimes I_d)\rho_i]
\eeq
 Note that the current operator, as measured in the STM experiment, involves \cite{bh} $\sigma_+\otimes s_z$ rather than $\sigma_+\otimes \mathbbm{1}$. This has a minor effect on the numerical data below.

\begin{figure}[t]
\includegraphics*[height=.3\columnwidth]{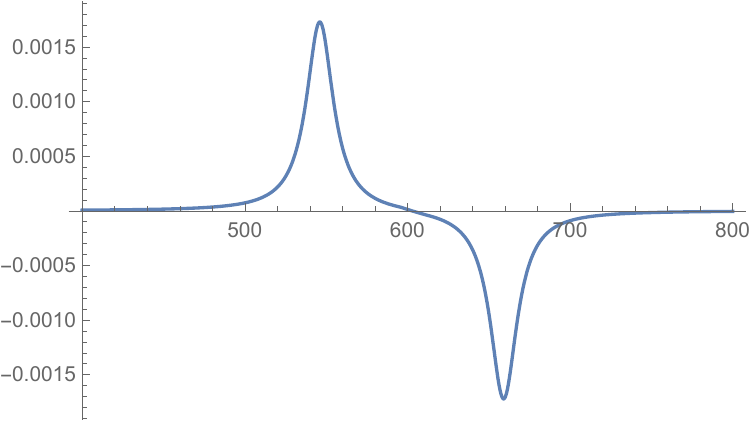}
\caption{{\bf projected ESR.} ESR resonance correction $\bar C_3(\nu)$ due to $\rho_3$. Parameters are $\nu_1=600,\,\nu_2=650,\,\nu_n=1,\,a=20,\,b=20,\,c=20,\,d=20,\,\gamma=0.1,\,\gamma_b=0.1,\,\gamma_1=10,\,\gamma_2=10, g=0, V=10^9$ ($V=0$ gives the same result).}
\label{ESR}
\end{figure}

For the parameters of Fig. \ref{ESR} we find that the 3rd eigenvalue satisfies the required constraints, in particular it has a phase that corresponds to $\nu_n$ and a reasonable overlap $\langle \tilde\tau\rho_3\rangle=-0.02217$. Fig. \ref{ESR} is for the correlation $\bar C_3(\omega)$ using the 3rd eigenvector, showing as expected an increase of one hyperfine state and a decrease of the other. When using the 1st eigenvector $C_1(\omega)$ has 2 equal peaks, the ratio $C_3(\omega)/C_1(\omega)$ at maximum varies from .007 at $\gamma=.01$ to .063 at $\gamma=1$; this ratio is well correlated with $\langle\tilde\tau\rho_3\rangle$ with the ratio  $\langle\tilde\tau\rho_3\rangle/\frac{C_3(\omega)}{C_1(\omega)}$ being 1.74-1.56 in this range.

For the parameters of Fig. \ref{ESR} the weight of the $|20\rangle$ states (within the staedy state $\rho_1$) is 0.2, i.e. all 10 states have equal weight. If the $|20\rangle$ state is neutral (e.g. the toluene case) then it is 80\% ionized.

\begin{figure}[t]
\includegraphics*[height=.28\columnwidth]{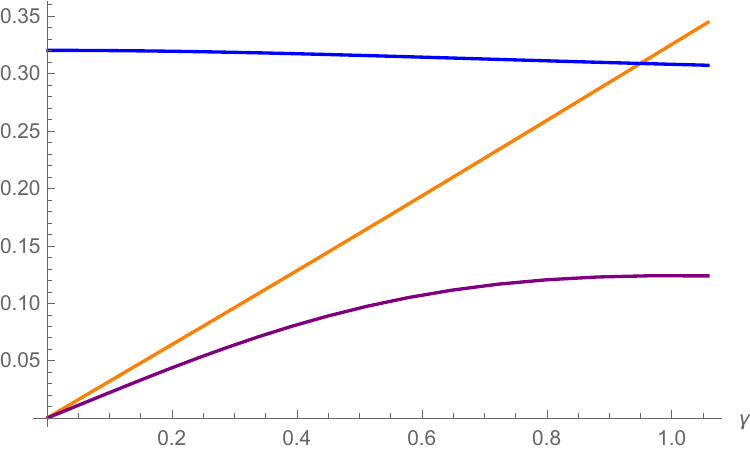}
\caption{{\bf Dependence on relaxation rates.} Dependence on $\gamma=\gamma_b$ of the frequency (phase of $\lambda_3/\bar T$, blue), linewidth ($\Gamma=-\ln|\lambda_3|/\bar T$, orange) and the projection $\tr[\tilde\tau\rho_3]$ (purple). Other parameters are same as in Fig. \ref{ESR}. }
\label{gamma dependence}
\end{figure}

It is instructive to look at the $\gamma$ dependence of $\lambda_3$ and of $\tilde\tau\rho_3$, see Fig. \ref{gamma dependence}. In particular if $\gamma=\gamma_b$ both vanish then also the linewidth $\Gamma=-(\ln(|\lambda_3|)/\bar T $ as well as $\tilde\tau\rho_3$ vanish. Since we need a small $\Gamma$ and a large $\tr[\tilde\tau\rho_3]$ we need a small $\gamma$, but not too small. When $\gamma\rightarrow 0$ but $\gamma_b$ stays finite then $\Gamma$ also stays finite.

 We consider briefly a regular NMR experiment, i.e. the correlation of $\tilde\tau$ in the full 10 dimensional phase space (our NMR-STM experiment is considered in the next subsection).
Consider $\gamma=\gamma_b$ so that the system is in steady state and we can use using the regression theorem, similar to Eq. \eqref{e14}, operating in both $|11\rangle$ and $|20\rangle$ spaces in the steady state $\rho_1$, i.e.
\beq{141}
C^{regular}_{NMR}(\nu)=\tr[(\tilde\tau\otimes I_d)\frac{1}{-i\omega-L_1}(\tilde\tau\otimes I_d)\rho_1]
\eeq

\begin{figure}[tb]
\includegraphics*[height=.3\columnwidth]{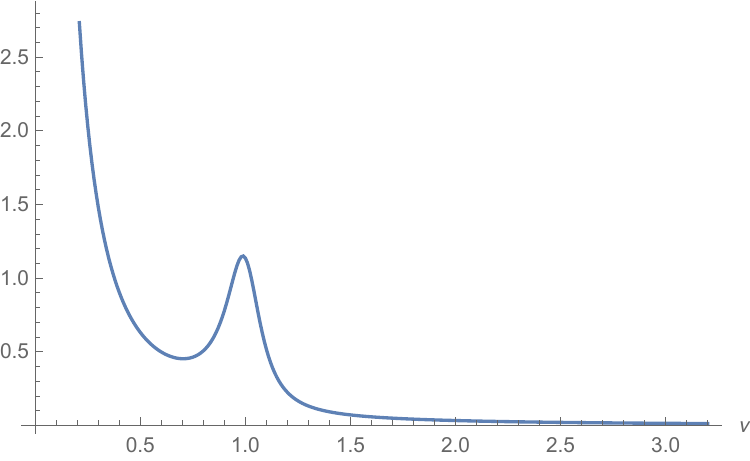}
\includegraphics*[height=.3\columnwidth]{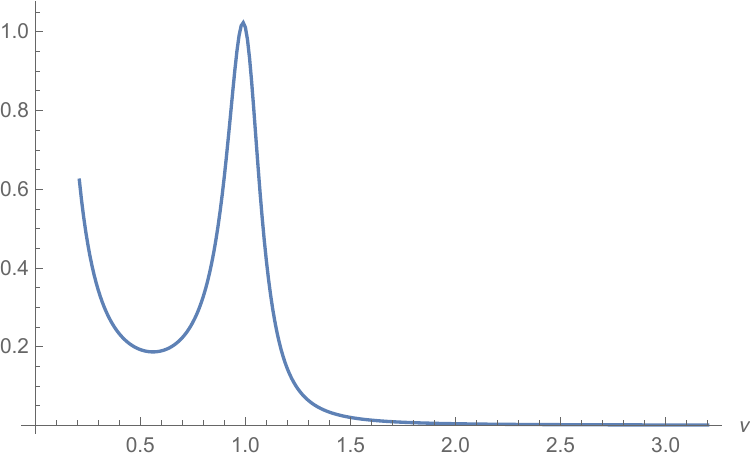}
\caption{{\bf Regular nuclear correlation.} Nuclear spin correlation $\langle\tilde\tau(t)\tilde\tau(0)\rangle_\nu$ using Eq. \eqref{e141}. Left:  $\nu_1=600,\,\nu_2=650,\,\nu_n=1,\,a=20,\,b=20,\,c=20,\,d=20,\,\gamma=0.1,\,\gamma_b=0.1,\,\gamma_1=10,\,\gamma_2=10, g=0, V=0$. Right: same parameters except $b=c=0$.}
\label{fullNMR}
\end{figure}

This nuclear correlation is plotted in Fig. \ref{fullNMR}. There is a strong peak at $\nu=0$ and a peak at the expected $\nu_n=1$ with a linewidth as expected for $\rho_3$ with $|\lambda_3|=0.968524=\eexp{-0.1\bar T}$, i.e. linewidth of 0.1. This correlation can be evaluated for the general case $L_1\neq L_2$, the method is developed in the next subsection.

\subsection{NMR-STM}

We arrive now at our main goal of simulating our NMR-STM experiment; consider first a qualitative argument.
 Asssume that the hyperfine peaks vary slowly with the instantaneous value of $\tilde\tau$, at a frequency $\sim\nu_n$. For small dissipation $\gamma,\,\gamma_b$ these peaks are given by eigenvalues of $\half\nu_1\sigma_z+\tilde a\sigma_z\otimes\tilde\tau$ where $\tilde a=\sqrt{a^2+d^2}$, i.e. ESR lines at $\nu_1\pm 2\tilde a$. The instantaneous probability of having nuclear spin up is $\half(1+\tilde\tau)$ determines the relative intensity of the $\nu_1+ 2\sqrt{a^2+d^2}$ line, similarly $\half(1-\tilde\tau)$ for the lower hyperfine line. Since the hyperfine peak is measured only in the subspace of $|11\rangle$ the correlation of $\tilde\tau$  has to be projected on $|11\rangle$ within the steady state $\rho_1$, this projection results in a significant reduction of the NMR signal.

As a somewhat more quantitative argument we consider the ESR Eq. (\ref{e14}) with weak relaxation $\gamma_1$. Within the $|11\rangle$ subspace it has the form
\beq{18}
C_1(\omega)&&=\int_{-\infty}^\infty \eexp{i(\half\nu_1+\tilde a\tilde\tau)\sigma_z t}\sigma_-
\eexp{-i(\half\nu_1+\tilde a\tilde\tau)\sigma_z t}\sigma_+\eexp{i\omega t-\gamma_1|t|}dt=
\int_{-\infty}^\infty \eexp{i(-\nu_1-2\tilde a\tilde\tau + \omega) t-\gamma_1|t|}\sigma_-\sigma_+ dt
\nonumber\\&& =\frac{2\gamma_1}{(\nu_1+2\tilde a\tilde\tau-\omega)^2+\gamma_1^2}\sigma_-\sigma_+\approx [\frac{1}{\gamma_1}(\mathbbm{1}+\tilde\tau)
+\frac{\gamma_1}{4\tilde a^2}(\mathbbm{1}-\tilde\tau)]\sigma_-\sigma_+
\eeq
where in the 2nd line we measure one hyperfine line with $\omega\approx\nu_1+2\tilde a$. Note that if we were to include triplet excited states within $|20\rangle$ that have a high energy $E_\infty$ then $\sigma_-\sigma_+|20\rangle$ would have a resonance at $\sim E_\infty$ so that its correlation is negligible at $\nu_1+2\tilde a$.

Thus, we define an NMR-STM operator for $\tilde a\gg \gamma_1$ 
\beq{19}
\bar \tau=(\mathbbm{1}-\sigma_z)\otimes\mathbbm{1}\otimes\tilde\tau\oplus 0\cdot|20\rangle\langle 20|
\eeq
The $1-\sigma_z$ factor projects on $\omega>0$ while $\tilde\tau$ projects on hyperefine states with opposite signs. For just the upper transition replace $\tilde\tau\rightarrow 1+(\tilde\tau)/2$ corresponding to Eq. \ref{e18}.   

\begin{figure}[t]
\includegraphics*[height=.3\columnwidth]{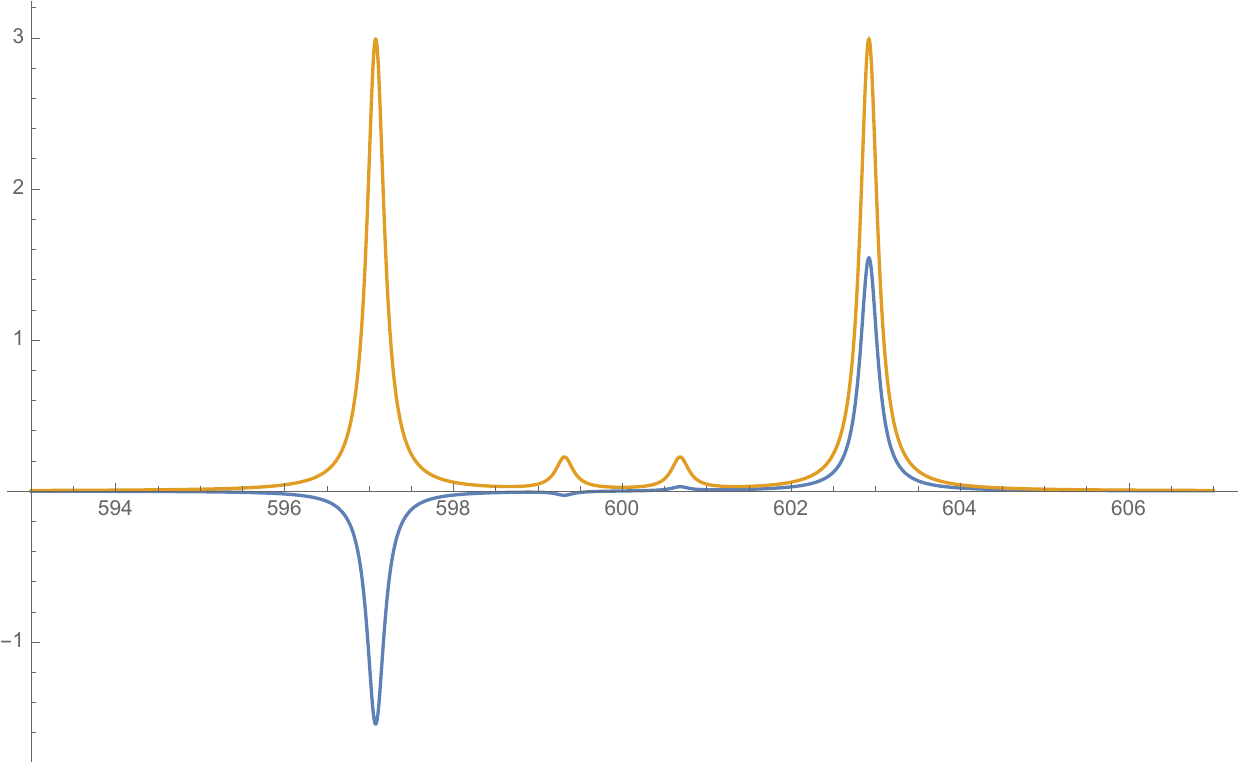}
\caption{{\bf Correlation of nuclear operator with ESR.} The correlation function $\langle \hat C(\omega) \bar \tau \rangle$ (blue) and the ESR spectrum $C(\omega)$ (red, enhanced by factor 2 for clarity) as functions of $\omega$. Parameters are $\nu_1=600,\,\nu_2=650,\,\nu_n=1,\,a=1.0,\,b=0.0,\,c=0.0,\,d=1.0,\,\gamma=0.1,\,\gamma_b=0.1,\,\gamma_1=0.1,\,\gamma_2=0.1.$ We observe the perfect (anti) correlation.}
\label{ESRoverlap}
\end{figure}

To further motivate the use of $\tilde\tau$ to represent our NMR-STM data we consider now the full form of the measured ESR correlation which is
\be
C(\omega)=\int_0^\infty\langle\sigma_-(t)\sigma_+(0)\rangle\eexp{i\omega t}dt+
c.c.
\ee
The regression theorem then gives the steady state of Eq. \eqref{e14}
\be
C(\omega)=2 {\rm Re}{\rm Tr}\left[ (\sigma_-\otimes I_d) \,\frac{1}{-i\omega - L}\,(\sigma_+\otimes I_d)\rho_{1}\right]\ .
\ee
This quantity is measured many times near a certain value of $\omega$ (with some bandwidth) and the results are Fourier transformed in order to obtain the power spectrum of the NMR signal. We want to demonstrate that measuring $C(\omega)$ near the frequency corresponding to one of the hyperfine lines is equivalent to measuring the appropriate projection of the nuclear spin, i.e. $\tilde \tau$. In order to convince the reader that this procedure is meaningful we introduce the following super-operator that represents the ESR operator
\be
\hat C(\omega) \equiv (\sigma_-\otimes I_d) \,\frac{1}{-i\omega - L}\,(\sigma_+\otimes I_d) + h.c.
\ee
and calculate the overlap between this ESR operator and $\bar\tau$,
\be
\langle \hat C(\omega) \tilde \tau \rangle = {\rm Tr} \left[\hat C(\omega) (\bar \tau \otimes I_d) \rho_{1}\right]\ .
\label{nuclearESR}
\ee
The results are shown in Fig.~\ref{ESRoverlap}. We observe the perfect correlation (anti-correlation) with the upper (lower) hyperfine line which justifies
using the operator $\tilde \tau$ to represent one of the ESR spectral lines.
(The weak signals in Fig.~\ref{ESRoverlap} are additional ESR transitions, visible when $\nu_n$ is comparable to $a,\, d$.)

We next derive the nuclear correlation for the general case $L_1\neq L_2$ using $\bar\tau$ of Eq. \eqref{e19}. In the supermatrix notation the density matrices $\rho_i$ in Eq. \eqref{e11} are vectors (of length 100). Assuming that these vectors span the whole space (corresponding to invertible $E$) then $\bar\tau\cdot\rho_k=\sum_n d_{kn}\rho_n$. The correlation, using the regression theorem for $N_1>N_2$ is
\beq{15}
&&C_{nuc}(N_1-N_2)=\langle \bar\tau(N_1)\bar\tau(N_2)\rangle=\tr[\bar\tau U^{N_1-N_2}\bar\tau\rho_1]=
\sum_n d_{1n}\lambda_n^{N_1-N_2}\tr[\bar\tau\rho_n]=\sum_n d_{1n}d_{n1}\lambda_n^{N_1-N_2}\nonumber\\
C_{nuc}(\nu)&&=\sum_n d_{1n}d_{n1}\int_0^\infty \eexp{-\Gamma_n t-i\nu_n^* t +i\nu t}dt+c.c.=\sum_n d_{1n}d_{n1}\frac{1}{-i(\nu-\nu_n^*)+\Gamma_n}+c.c.
\eeq
where $\tr[\rho_n]=\delta_{n,0}$, the eigenvalues $\lambda_n=|\lambda_n|\eexp{i\varphi_n}$ define $\Gamma_n=-(\ln|\lambda_n|)/\bar T,\,\nu^*_n=\varphi_n/\bar T$ and the c.c. comes from summation on $N_1<N_2$. Using the previously defined matrix $E$ of eigenvectors $E_{ij}=(\rho_j)_i$ we obtain for the matrix $\hat d$, $(\hat d)_{kn}=d_{kn}$
\beq{16}
(\tilde\tau\cdot E)_{ik}&&=\sum_j(\tilde\tau)_{ij}(\rho_k)_j=\sum_n d_{kn}(\rho_n)_i=\sum_n d_{kn}E_{in}
=(E\cdot\hat d^{T})_{ik}\nonumber\\
\Rightarrow\qquad \hat d&&=(E^{-1}\cdot \tilde\tau\cdot E)^{T}
\eeq
which is an efficient way of evaluating $d_{kn}$. Thus finally
\beq{20}
 C_{nuc}(\nu)=\sum_{n>0}\re[ \frac{2d_{1n}d_{n1}}{-i(\nu-\nu_n^*)+\Gamma_n}]
\eeq
For small $\gamma,\gamma_d$ we find a weak signal at $\nu_n=1$MHz, while if the hyperfine couplings $a,d$ are small (and $\gamma_1,\,\gamma_2$ are corresponding small as they need to be smaller than $a,\, d$) we find a large signal at a shifted position, see Fig. 5a.

To interpret the shifted signal we note that
when $\nu_n$ is comparable to $a,\,d$ and $\gamma_1\ll a,\,d$ then the nucleus feels an effective magnetic field corresponding to a Hamiltonian $(\half\nu_n\mathbbm{1}+a\sigma_z)\otimes\tau_z+d\sigma_z\tau_y$. The shifted frequency is then the eigenvalue difference $\tilde\nu_n=\sqrt{(\nu_n+2a)^2+(2d)^2}$ taking $\sigma_z\rightarrow +1$. We consider this as an extreme case of a chemical shift, i.e. a shifted NMR due to electron polarization. We note that the more conventional chemical shift is obtained  when $\sigma_z$ is averaged, i.e. polarization effect \cite{wertz}. In fact if $\gamma_1$ increases then eventually $\langle\sigma_z\rangle$ vanishes and the peak approaches $\nu_n$ (for $\gamma_1\gtrsim 20$). One can maintain a finite $\langle\sigma_z\rangle$ by keeping different relaxations of the up and down electron spin, leading to a chemical shift even for strong relaxations. We note that the chemically shifted peak persists also at higher hyperfine couplings, though with reduced intensity.

It is interesting to consider the situation at large $\gamma_1>2\tilde a$, which is a type of Zeno effect. Although the hyperfine lines merge the correlation Eq. \eqref{nuclearESR} changes sign at $\nu_1$ so that $\tilde\tau$ can still be used to measure NMR. Furthermore, we find that the NMR signal becomes sharper as $\gamma_1$ increases. To motivate this remarkable effect we note that the evolution of the raising operator $\tau_+$ for the eigenstates of $\tilde\tau$ is, neglecting here $\nu_n$,
$\tau_+(t)=\eexp{-i\tilde a\tilde\tau\int_0^t \sigma_z(t')dt'}\tau_+\eexp{i\tilde a\tilde\tau\int_0^t\sigma_z(t')dt'}=\eexp{-2i\tilde a \int_0^t\sigma_z(t')dt'}\tau_+$, hence the average $\langle\tau_+(t)\rangle\sim\eexp{-\gamma_\varphi t}$ where the decay rate is $\gamma_\varphi =2\tilde a^2/\gamma_1$ using the $\sigma_z$ correlation at $\omega=0$ which is $1/\gamma_1$.

\begin{figure}[tb]
\includegraphics*[height=.3\columnwidth]{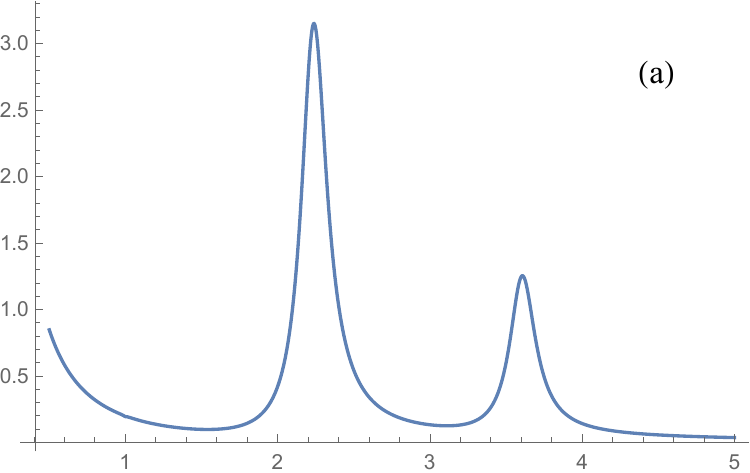}
\includegraphics*[height=.3\columnwidth]{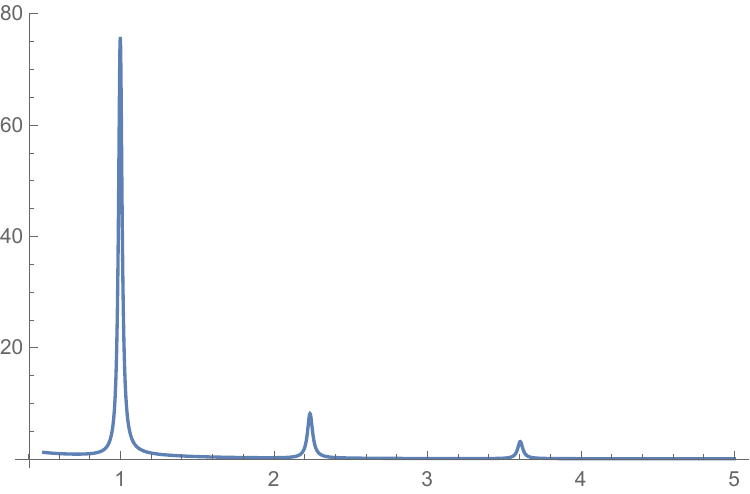}
\caption{{\bf NMR-STM with both $\pm\omega$.} NMR-STM spectra when both $\pm\omega$ ESR resonances are summed. Left: $\gamma=0.01,\, \gamma_1=0.1,\,a= d=1$. Right: $\gamma=10,000,\,\gamma_1=0.01,\,a=d=1$. Compare with Fig. 5 of the main text where only one chemically shifted peak is seen, corresponding to $\omega>0$ ESR.}
\label{2peak}
\end{figure}

In some cases measurements sum both $\pm \omega$ ESR resonances, hence one needs the combination $2(\sigma_-\sigma_+ + \sigma_+\sigma_-)=2\mathbbm{1}$. The NMR-STM operator is then
\beq{21}
\bar{ \bar\tau}=2\cdot\mathbbm{1}\otimes\mathbbm{1}\otimes\tilde\tau\oplus 0\cdot|20\rangle\langle 20|
\eeq
The corresponding spectrum is shown in figure \ref{2peak}. The 2 peaks correspond now to both $\pm 1$ electron spin, hence $\tilde\nu_n=\sqrt{(\nu_n\pm2a)^2+(2d)^2}$, displaying two chemically shifted resonances.

Since we do not know at present the relaxation rates $\gamma,\,\gamma_b$, we have considered also large values, though still small on the voltage scale (the chemical potential difference of the two molecules is $\sim 1$V). We have found that sharp NMR signals are possible also in this case. We note that in this case the singlet states rapidly decay while the $\gamma_1,\,\gamma_2$ transitions populate the triplet states, leading to dominance of the latter, hence a strong coupling between the two spins. Projecting the Hamiltonian (keeping only the $a,\, d$ hyperfine terms) onto the triplet states we find that the ESR frequencies are
\beq{21}
\nu_\pm^{ESR}=\half[\nu_1+\nu_2\pm\nu_n\pm\sqrt{4(a^2+d^2)-4a\nu_n+\nu_n^2}]
\eeq
These frequencies can be detected by measuring either spin. We find that a sharp NMR frequency at $\nu_n$ is present when $\gamma_1\,\gamma_2$ are small and if also $a,\,d$ are small then two signals appear, one at $\nu_n$ and the other at a shifted value, given by the same $\tilde\nu_n$ as above, see the bottom figures of Fig. 5. In case that both $\pm\omega$ ESR are summed there are two chemically shifted resonances as shown in Fig. \ref{2peak} (right).

\subsection{NMR-STM with two nuclear spins}

In almost all molecules there are distinct nuclei with various hyperfine couplings. In particular in toluene there are inequivalent $^1$H nuclei with either strong or weak hyperfine coupling \cite{wertz}. Since the ESR (Fig. 4a) shows fairly strong hyperfine splitting of $\sim$10MHz, while interesting structure is manifested by weakly coupled nuclei (Figs. 5, \ref{2peak}) we are motivated to study the case of two nuclei spins, one spin with a strong hyperfine coupling, producing the dominant structure in the ESR, while the other with a weak coupling. The latter may show splitting of the dominant ESR lines, however, when its hyperfine coupling is weaker than the linewidth then it is not even noticeable in the ESR. As we find here, the correlation of the latter nuclear spin produces sharp structure at the nuclear frequency that is sharper as the ESR linewidth increases. We propose that this situation is the best scenario accounting for our NMR-STM data.

\begin{figure}[tb]
\includegraphics*[height=.15\columnwidth]{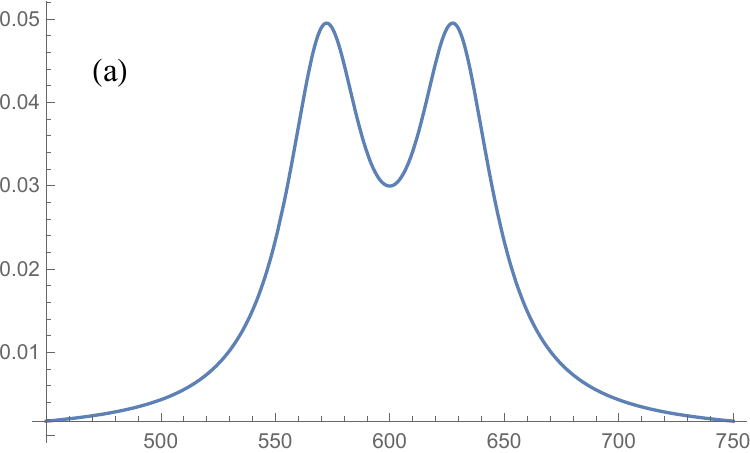}
\includegraphics*[height=.15\columnwidth]{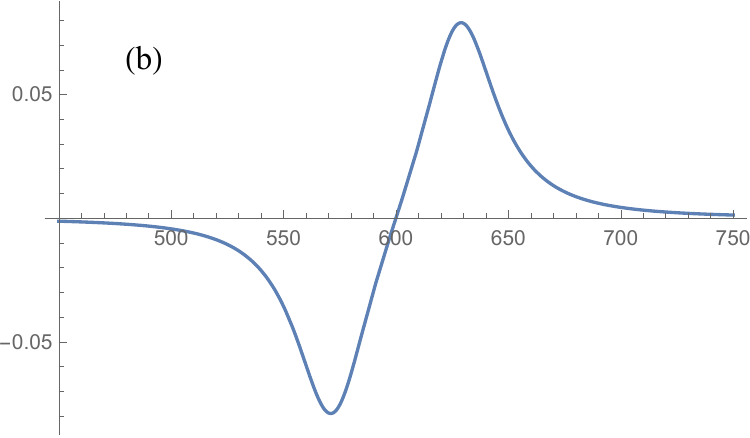}
\includegraphics*[height=.15\columnwidth]{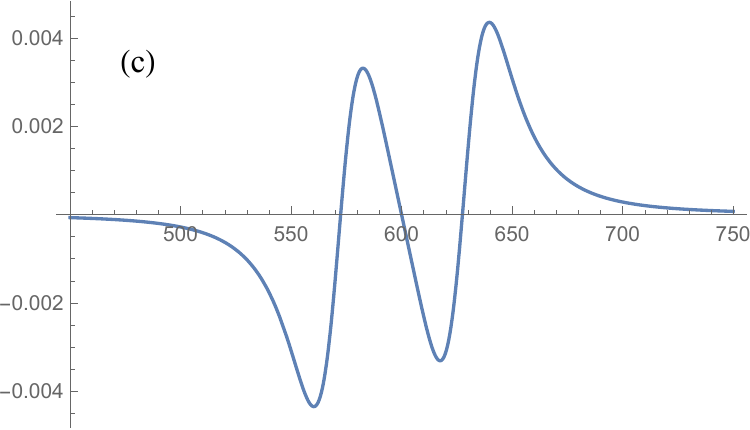}
\caption{{\bf Case of 2 nuclei}. (a) ESR spectra, (b) overlap of the ESR operator with $\tau'$ of Eq. \eqref{e23}, (c) overlap of the ESR operator with $\rho'$. Parameters in the Hamiltonian Eq. \eqref{e22}: $\nu_1=600,\, \nu_2=650,\,\nu_{n1}=\nu_{n2}=1,\,\gamma=0.1,\,\gamma_1=20,\, a=d=10,\, a_2=d_2=0.5$}
\label{2nuclei}
\end{figure}

We consider then a Hamiltonian of four spins so that the $|11\rangle$ state has 16 states while $|20\rangle$ has 4 states, i.e. our Hilbert space has now 20 states; the weakly coupled spin corresponds to Pauli matrices $\rho_x,\,\rho_y,\,\rho_z$.
\beq{22}
&&{\cal H}_{11}=[\half\nu_1\sigma_z\otimes\mathbbm{1}\otimes\mathbbm{1} +\half\nu_2 \mathbbm{1}\otimes s_z \otimes \mathbbm{1} +\half\nu_n\mathbbm{1}\otimes\mathbbm{1}\otimes\tau_z+a\sigma_z\otimes\mathbbm{1}
\otimes\tau_z +d\sigma_z\otimes\mathbbm{1}\otimes\tau_y]\otimes \mathbbm{1} \nonumber\\&&
\qquad\qquad+\half\nu_{n2}\mathbbm{1}\otimes\mathbbm{1}\otimes\mathbbm{1}\otimes\rho_z+a_2\sigma_z\otimes\mathbbm{1}
\otimes\mathbbm{1}\otimes\rho_z+d_2\sigma_z\otimes\mathbbm{1}\otimes\mathbbm{1}\otimes\rho_y\nonumber\\&&
{\cal H}_{20}=\half\nu_n\tau_z\otimes\mathbbm{1}+\half\nu_{n2}\mathbbm{1}\otimes\rho_z\nonumber\\&&
{\cal H}={\cal H}_{11}|11\rangle\langle 11|+{\cal H}_{20}|20\rangle\langle 20|
\eeq
where we focus on the more relevant hyperfine couplings to $\tau_z,\,\tau_y,\,\rho_z,\,\rho_y$. Our results for the ESR spectra are shown in Fig. \ref{2nuclei}a. The linewidth and $a,d$ are chosen to correspond to the observed ESR shape, Figs. S2a, S2b, the weaker hyperfine $a_2=d_2=.5$ is not seen since it is smaller than the linewidth. To study the nuclear correlations we define the operators 
\beq{23}
&&\tau'=2\cdot\mathbbm{1}\otimes\mathbbm{1}\otimes\tilde\tau\otimes\mathbbm{1}\oplus 0\cdot|20\rangle\langle 20|\nonumber\\&&
\rho'=2\cdot\mathbbm{1}\otimes\mathbbm{1}\otimes\mathbbm{1}\otimes\tilde\rho\oplus 0\cdot|20\rangle\langle 20|
\eeq
where $\tilde\rho=(a_2\rho_z+d_2\rho_y)/\tilde a_2$ and $\tilde a_2=\sqrt{a_2^2+d_2^2}$.
The overlap of the ESR spectra with the first nucleus, using Eq. \eqref{nuclearESR} with $\bar\tau\rightarrow\tau'$, is shown in Fig. \ref{2nuclei}b, it is insensitive to to second nucleus as expected and is qualitatively similar to Fig. \ref{ESRoverlap}. The ESR overlap with the second nucleus, using Eq. \eqref{nuclearESR} with $\bar\tau\rightarrow\rho'$,  is shown in Fig. \ref{2nuclei}c. It shows clearly the splitting of the ESR spectrum to four lines (using $\mathbbm{1}+\tilde\rho$ would show only two lines). We show the nuclear correlations for $\tau'$ and $\rho'$ using the formulation as in Eq. \eqref{e20}  in Fig. 5b of the main text (we do not show the $\mathbbm{1}+\tau'$ or $\mathbbm{1}+\rho'$ correlations to conform with Figs. \ref{2nuclei}b,\ref{2nuclei}c, the NMR spectra are very similar).
To measure the NMR signal the sampling of the ESR data should be taken in the vicinity of one of the peaks (or one of the dips) in Fig. \ref{2nuclei}c. This vicinity has a reasonable overlap with $\rho'$, it is smaller than the overlap with $\tau'$ yet its NMR signal is considerably stronger,  thus the $\rho'$ correlation is a proper presentation of the NMR-STM data with this sampling.

 The results in Fig. 5b are remarkable: the 1st spin shows a very weak and negative dip at $\nu=1$ while the 2nd spin shows a strong and sharp signal at  $\nu=1$ while the ESR spectra has a linewidth in agreement with the data. The NMR linewidth varies as $\tilde a_2^2/\gamma_1$, similar to the Zeno effect discussed above, though now there is just one frequency at $\nu=\nu_{n2}$. We therefore consider this case with two nuclear spins as the most likely to account for our data.
